\documentclass[journal]{IEEEtran}
\AtBeginDocument{\ifdefined\proof\else\newenvironment{proof}{\begin{IEEEproof}}{\end{IEEEproof}}\fi}

\usepackage{cite}
\usepackage{amsmath,amssymb,amsfonts}
\usepackage{bm,mathtools,mathrsfs}
\usepackage{graphicx}
\usepackage{booktabs}
\usepackage{algorithm}
\usepackage{algpseudocode}
\usepackage{comment}
\usepackage{writingkit}

\setboolean{showModification}{false}
\setboolean{showRevA}{false} %
\setboolean{showRevB}{true} %

\usepackage[
    colorlinks,
    linkcolor=blue,
    anchorcolor=blue,
    citecolor=blue, 
    hyperfigures=true,
    urlcolor=black,
]{hyperref}

\graphicspath{{figures/}}
\algrenewcommand\algorithmicrequire{\textbf{Input:}}

\newcommand{\R}{\mathbb{R}}

\newcommand{\B}{\mathbb{B}}
\newcommand{\col}{\operatorname{col}}
\newcommand{\rank}{\operatorname{rank}}

\newcommand{\diag}{\operatorname{diag}}
\newcommand{\T}{\mathsf T}
\newcommand{\Tini}{T_{\mathrm{ini}}}
\newcommand{\Td}{T_{\mathrm d}}
\newcommand{\Up}{U_{\mathrm p}}
\newcommand{\Uf}{U_{\mathrm f}}
\newcommand{\Yp}{Y_{\mathrm p}}
\newcommand{\Yf}{Y_{\mathrm f}}
\newcommand{\Wp}{W_{\mathrm p}}
\newcommand{\xh}{\hat x}
\newcommand{\Xh}{\widehat X}
\newcommand{\Ez}{\widehat E_z}
\newcommand{\Th}{\Theta}
\newcommand{\Ys}{\mathcal Y_s}      %
\newcommand{\U}{\mathcal U}
\newcommand{\Y}{\mathcal Y}
\newcommand{\Lyap}{\mathscr W}
\newcommand{\nrm}[1]{\lVert #1\rVert}
\newcommand{\wnrm}[2]{\lVert #1\rVert_{#2}^{2}}

\newcommand{\ee}{\bar e}
\newcommand{\dd}{\bar d}
\newcommand{\smin}{\sigma_{\min}}

\newcommand{\calA}{\mathcal A}
\newcommand{\calW}{\mathcal W}
\newcommand{\Ls}{\mathscr L}
\newcommand{\Ss}{\mathscr S}
\newcommand{\J}{\mathcal J}
\newcommand{\proofhead}[1]{\par\textit{#1:}\enspace}

\newtheorem{problem}{\textbf{Problem}}
\newtheorem{theorem}{\textbf{Theorem}}
\newtheorem{corollary}{\textbf{Corollary}}
\newtheorem{lemma}{\textbf{Lemma}}
\newtheorem{remark}{\textbf{Remark}}
\newtheorem{definition}{\textbf{Definition}}
\newtheorem{assumption}{\textbf{Assumption}}

\usepackage{booktabs}

\newcommand{\PaperAuthors}{Jingyuan Li, Dawei Shi, and Ling Shi}

\newcommand{\PaperThanks}{%
\thanks{This work has been submitted to the IEEE for possible publication. Copyright may be transferred without notice, after which this version may no longer be accessible.}
\thanks{J. Li and D. Shi are with the MIIT Key Laboratory of Servo Motion System Drive and Control, School of Automation, Beijing Institute of Technology, Beijing 100081, China (e-mail: jingyuan.li@bit.edu.cn, daweishi@bit.edu.cn).}
\thanks{L. Shi is with the Department of Electronic and Computer Engineering and the Department of Chemical and Biological Engineering, Hong Kong University of Science and Technology, Clear Water Bay, Hong Kong SAR. (e-mail: eesling@ust.hk).}
}

\newcommand{\PaperBios}{}

\begin{document}

\title{DeoPC: Data-Enabled Output-Feedback Predictive Control for State Regulation}

\author{\PaperAuthors\PaperThanks}
\maketitle

\begin{abstract}
Constrained regulation of the internal state of an unknown linear system is challenging when only input--output measurements are available online.
In this work, we develop a data-enabled output-feedback predictive control (DeoPC) framework for constrained state regulation using input--output data.
Specifically, a data-driven state estimate is constructed from past input--output data via a subspace-based singular value decomposition.
Based on this state representation, we formulate a predictive controller that directly penalizes the state and control effort while enforcing the system constraints.
Under persistent excitation and initialization conditions, we establish recursive feasibility and closed-loop stability of DeoPC in both noise-free and bounded-noise settings.
In addition, we show that DeoPC can be calibrated to the physical state coordinates using a few offline state measurements, or treated as a model predictive control tuning problem when no state measurements are available.
The effectiveness of DeoPC is demonstrated through simulations on an unstable batch reactor system.
\end{abstract}


\section{Introduction}\label{sec:intro}

Recent years have seen increasing attention to data-driven control~\cite{hou2013model,markovsky2021behavioral}.
Among various approaches, data-enabled predictive control (DeePC)~\cite{coulson2019data} has emerged as a popular method due to its simplicity and effectiveness, and has found applications in a wide range of fields including quadcopters~\cite{elokda2021data}, power systems~\cite{huang2022decentralized}, synchronous motor drives~\cite{carlet2022data}, and building climate control~\cite{chinde2022data}.
DeePC typically minimizes output-tracking errors through a constrained receding-horizon optimization problem, using input--output data to predict future system trajectories.
However, many control tasks require tracking unmeasurable internal states, which poses a challenge for data-driven control based solely on input--output data~\cite{joseph1978inferential,tsuji2026predictive}.

Several lines of data-driven predictive control research are relevant to this problem.
Data-driven linear-quadratic regulation~\cite{depersis2020formulas,dorfler2023certainty} computes a state-feedback law for a prescribed quadratic state cost directly from input--state data, but requires full-state measurements both offline and online.
Another line of work constructs state representations from past inputs and outputs, as in subspace linear-quadratic-Gaussian design~\cite{favoreel1999model,dong2008equivalence}, output-feedback regulation~\cite{dai2023data,alsalti2024efficient}, and non-minimal state-space model predictive control (MPC)~\cite{zhang2011improved,zhang2018systematic}.
In these designs, the performance index is specified on outputs and inputs or on their history, so an individual internal state cannot be weighted.
A third line of work identifies a state-space model from input--output data and places the cost on its state~\cite{wang2022modeling,balim2024data}. 
However, the guarantees in~\cite{wang2022modeling} hold for the identified model and neglect the identification error, while those in~\cite{balim2024data} account for it but rely on prior knowledge of the model structure.

Regarding the closed-loop guarantees for this problem, the existing results for constrained data-driven output feedback have been developed mainly for DeePC.
Berberich et al.~\cite{berberich2021data} used a terminal equality constraint to prove recursive feasibility and exponential stability of DeePC, as well as practical exponential stability when the output data are corrupted by bounded noise.
In subsequent developments, terminal ingredients were constructed from data~\cite{berberich2021terminal,yang2026unified}, a sufficiently long horizon was shown to make them unnecessary~\cite{bongard2023robust}, noisy data were handled via a truncated singular-value decomposition~(SVD)~\cite{alsalti2026robust}, and a window of past inputs and outputs was used as an extended state for feedback and constraint tightening in a robust scheme~\cite{kloppelt2025novel}.
In these results, the Lyapunov function is constructed from the optimal cost and an input--output-to-state stability (IOSS) storage function. This storage function decreases up to terms in the input and output, which are directly bounded by the input--output stage cost.
However, when the stage cost is formulated in terms of a state built from input--output data, it no longer bounds these terms in general.
The storage function should then be matched with the state weight, and its existence should be certified from input--output data, even when the data are noisy.
The above studies leave open the problem of how to directly penalize the unmeasurable internal states in a data-driven predictive control framework using only input--output data, while accounting for noise and ensuring closed-loop guarantees.
In this paper, we develop a new data-enabled output-feedback predictive control (DeoPC) framework to address this problem.
Inspired by the subspace-based construction~\cite{favoreel1999spc,vanoverschee1996subspace}, we construct from past input--output data a data-driven state estimate that equals the internal state up to an invertible transformation in the noise-free case, and we formulate DeoPC by directly penalizing this state and the input while enforcing input and output constraints.
However, several challenges need to be addressed to enable our design, including how a data-driven state estimate  can be constructed from input--output data to enable state-weighted predictive control, how recursive feasibility and closed-loop stability can be guaranteed in both noise-free and bounded-noise settings, and how the proposed DeoPC framework can be implemented in practice, with or without access to offline state measurements.
The main contributions of the developed results are summarized as follows:
\begin{enumerate}
\item The DeoPC framework is proposed for constrained predictive control using input--output data.
Unlike existing input--output data-driven predictive control frameworks~\cite{coulson2019data,berberich2021data} that specify the cost in terms of the outputs, DeoPC directly penalizes the internal state to enable flexible state regulation.
Specifically, a data-driven state estimate is constructed through an oblique projection and a truncated SVD which is then incorporated into the predictive controller with a quadratic cost on this state estimate~(\textbf{Algorithm~\ref{alg:deopc}}).
For noise-free and sufficiently exciting data, we show that DeoPC is equivalent to a model-based state-feedback MPC, in the sense that the two problems have the same feasible input sequences, optimal value, and optimal input sequence~(\textbf{Theorem~\ref{prop:equiv}}).

\item Recursive feasibility and closed-loop stability of DeoPC are established in both the noise-free and bounded-noise settings.
In contrast to existing stability results~\cite{berberich2021data,kloppelt2025novel}, in which the IOSS storage function is defined with respect to the outputs, we construct from data a quadratic storage function matched to the chosen state weight.
For noise-free data, we prove that the closed-loop trajectory satisfies the constraints and the data-driven state estimate converges exponentially to the origin~(\textbf{Theorem~\ref{thm:nominal}}).
For data corrupted by bounded noise, we prove that, under persistent excitation and initialization conditions, the constraints remain satisfied and the data-driven state estimate converges exponentially to a neighborhood of the origin whose size is proportional to the noise level~(\textbf{Theorem~\ref{thm:noisy}}).

\item Two implementations of DeoPC and numerical validation examples are presented.
When a few offline state samples are available, the transformation between the data-driven state estimate and the physical state is computed by least squares from paired samples, so that the state weight and the reference are specified directly in physical coordinates (\textbf{Algorithm~\ref{alg:align}}).
Without state samples, DeoPC can be applied directly, and its weight is tuned as in MPC without loss of design freedom (\textbf{Corollary~\ref{cor:tuning}}).
Simulations on an unstable batch reactor validate the theoretical results.
\end{enumerate}

The rest of the paper is organized as follows.
Section~\ref{sec:prelim} reviews MPC and DeePC as preliminaries and formulates the control problem.
Section~\ref{sec:deopc} constructs the data-driven state estimate and develops the DeoPC framework.
Section~\ref{sec:main} analyzes the theoretical properties of DeoPC in both noise-free and noisy-data settings.
Section~\ref{sec:impl} discusses the practical implementation of DeoPC with and without offline state samples.
Section~\ref{sec:num} presents numerical results for an unstable batch reactor system, followed by the conclusions in Section~\ref{sec:concl}.
The Appendix provides proofs of technical lemmas.

\emph{Notation:} Let $\col(a_1,\ldots,a_s)$ denote the vertical stack of vectors or matrices, and write $z_{[a,b]}=\col(z_a,\ldots,z_b)$.
For a symmetric matrix, $M\succ0$ ($M\succeq0$) means positive (semi)definite.
Let $\nrm{\cdot}$ denote the Euclidean norm or the induced spectral norm, and let $\wnrm{x}{Q}=x^\T Qx$.
Further, $M^\dagger$ denotes the Moore--Penrose pseudoinverse, $M_r^\dagger$ denotes the pseudoinverse of the rank-$r$ truncated SVD of $M$, and $\sigma_i(M)$ denotes the $i$-th largest singular value.
The closed Euclidean unit ball is denoted by $\B$.
For a matrix $M$, $\ker M$, $\operatorname{im}M$, and $\sigma_{\min}(M)$ denote its kernel, image, and smallest singular value, and for a symmetric $M$, $\lambda_{\min}(M)$ and $\lambda_{\max}(M)$ denote its smallest and largest eigenvalues.
$I_n$ is the $n\times n$ identity matrix and $\otimes$ is the Kronecker product.

\section{Preliminaries and Problem Formulation}\label{sec:prelim}

Consider the discrete-time linear time-invariant system
\begin{equation}\label{eq:plant}
x_{k+1}=Ax_k+Bu_k+w_k,\quad y_k=Cx_k+Du_k+v_k,
\end{equation}
where $x_k\in \mathbb{R}^n$ is the state, $u_k \in \mathbb{R}^m$ is the input, $y_k\in \mathbb{R}^p$ is the measured output, and $w_k$ and $v_k$ are the process noise and the measurement noise, respectively.
The system matrices $(A,B,C,D)$ are \emph{unknown}, and the state $x_k$ is not measured; only the inputs and the outputs are available.
The inputs and outputs are subject to the constraints $u_k\in\U$ and $y_k\in\Y$ for all $k\geq0$, where $\U$ and $\Y$ are compact sets that contain the origin in their interiors.
Throughout the paper, it is assumed that $(A,B)$ is controllable, $(C,A)$ is observable, and the state dimension $n$ is known.
Let $\ell$ denote the \emph{lag} of the system, defined as $\ell = \min\{j\geq 1 : \rank \mathcal O_j=n\}$, where $\mathcal O_j=\col(C,CA,\ldots,CA^{j-1})$ is the extended observability matrix.

An offline input--output trajectory $(u^{\rm d}_k,y^{\rm d}_k)$, $k=0,\ldots,\Td-1$, of~\eqref{eq:plant} is available, where the superscript ${\rm d}$ marks the data collected in the offline phase.
These data, as well as the online inputs and outputs, are either noise-free, i.e., $w_k=0$ and $v_k=0$, or corrupted by noise that satisfies the following assumption.

\begin{assumption}\label{ass:bounded-noise}
The noise is bounded, i.e., there exist $\bar{w}\geq 0$ and $\bar v \geq 0$ such that $\left\lVert w_k \right\rVert \leq \bar{w} $ and $\left\lVert v_k \right\rVert \leq \bar v$ for all $k$.
\end{assumption}

\subsection{Model Predictive Control}\label{sec:mpc}
For the noise-free model, i.e., \eqref{eq:plant} with $w=v=0$, the standard state-feedback MPC~\cite{rawlings2017model} with prediction horizon $N$ solves the following problem at each time $k$:
\begin{equation}\label{eq:ideal}
\begin{aligned}
\min_{u}\ &\sum_{i=0}^{N-1}\big(\wnrm{x_{i|k}}{Q_x}+\wnrm{u_{i|k}}{R}\big)+\wnrm{x_{N|k}}{P_x}\\
\text{s.t. }&x_{i+1|k}=Ax_{i|k}+Bu_{i|k},\quad x_{0|k}=x_k,\\
&u_{i|k}\in\U,\quad Cx_{i|k}+Du_{i|k}\in\Y,\quad i=0,\ldots,N-1,
\end{aligned}
\end{equation}
and applies $u_k=u^*_{0|k}$, where $Q_x\succeq0$ and $R\succ0$ are the stage weights and $P_x\succ0$ is the terminal weight.

\subsection{Data-Enabled Predictive Control}\label{sec:deepc}
For a signal $f$ of length $\Td$ and a depth $j \leq \Td$, the depth-$j$ block Hankel matrix is denoted by
\begin{equation}
H_j(f)=\begin{bmatrix}f_0&f_1&\cdots&f_{\Td-j}\\ \vdots&\vdots&&\vdots\\ f_{j-1}&f_j&\cdots&f_{\Td-1}\end{bmatrix}.   
\end{equation}
Let $\Tini$ be the length of the past window and $N$ the prediction horizon of~\eqref{eq:ideal}.
Throughout the paper, these horizons are chosen with $\Tini\geq\ell$ and $N\geq\ell$.
We construct the depth-$L$ Hankel matrices $H_L(u^{\rm d})$ and $H_L(y^{\rm d})$ with $L = \Tini+N$, and we partition them as
\begin{equation}\label{eq:partition}
\begin{bmatrix}\Up\\ \Uf\end{bmatrix}=H_L(u^{\rm d}),\quad
\begin{bmatrix}\Yp\\ \Yf\end{bmatrix}=H_L(y^{\rm d}),\quad
\Wp=\begin{bmatrix}\Up\\ \Yp\end{bmatrix},
\end{equation}
where the first $\Tini$ block rows give the past blocks $\Up$ and $\Yp$, the last $N$ block rows give the future blocks $\Uf$ and $\Yf$, and $\Wp$ collects the past blocks.
The richness of the offline input is measured as follows.

\begin{definition}[$\sigma$-persistent excitation \cite{coulson2023quantitative,berberich2023quantitative}]\label{def:spe}
Let $\sigma>0$.
The input $u^{\rm d}$ is said to be $\sigma$-persistently exciting of order $j$ if $H_j(u^{\rm d})$ has at least as many columns as rows and $\sigma_{\min}\big(H_j(u^{\rm d})\big)\geq\sigma$.
\end{definition}

\begin{remark}
An input is persistently exciting of order $j$ in the usual sense \cite{willems2005note}, i.e., $H_j(u^{\rm d})$ has full row rank, if and only if it is $\sigma$-persistently exciting of order $j$ for some $\sigma>0$.
\end{remark}

Let $\xi_k=\col(u_{[k-\Tini,k-1]},y_{[k-\Tini,k-1]})$ denote the window of the last $\Tini$ inputs and outputs.
For noise-free data, the following lemma characterizes the input--output behavior of the system.

\begin{lemma}[Willems' fundamental lemma~\cite{willems2005note}]
Consider system~\eqref{eq:plant} with $w=v=0$.
If $u^{\rm d}$ is persistently exciting of order $L+n$, then every input--output trajectory of length $L$ is a linear combination of the columns of $\col(\Wp,\Uf,\Yf)$.
\end{lemma}

DeePC~\cite{coulson2019data} uses this lemma as its prediction model and solves
\begin{equation}\label{eq:deepc}
\begin{aligned}
\min_{g,u,y}\ &\sum_{i=0}^{N-1}\big(\wnrm{y_i}{S}+\wnrm{u_i}{R}\big)\\
\text{s.t. }&\col(\Wp,\Uf,\Yf)\,g=\col(\xi_k,u,y),\\
&u_i\in\U,\quad y_i\in\Y,\quad i=0,\ldots,N-1,
\end{aligned}
\end{equation}
with $u=\col(u_0,\ldots,u_{N-1})$, $y=\col(y_0,\ldots,y_{N-1})$, and $S\succ0$.
Since $\Tini\geq\ell$, the noise-free past window fixes the initial state and hence the future outputs $y$ for given future inputs $u$ \cite{coulson2019data}.
For noise-free data that satisfy the required excitation and initialization conditions, unregularized DeePC and subspace predictive control (SPC) \cite{favoreel1999spc} yield the same optimal input--output trajectories \cite{fiedler2021relationship}, and with matched costs and constraints they are equivalent to model-based MPC \cite{dorfler2023bridging}.

\subsection{Problem Formulation}\label{sec:problem}
The MPC~\eqref{eq:ideal} penalizes the state but requires an accurate system model and direct state measurements, whereas DeePC~\eqref{eq:deepc} needs only input--output data but penalizes only inputs and outputs.
We therefore seek a controller that combines the state-based cost of MPC with the data-driven framework of DeePC.

\begin{problem}\label{prob:state}
Using only the offline trajectory $(u^{\rm d},y^{\rm d})$ and the past window $\xi_k$, as DeePC~\eqref{eq:deepc} does, design a receding-horizon controller with the constraints $u_k\in\U$ and $y_k\in\Y$ which uses a stage cost of the same quadratic form as in MPC~\eqref{eq:ideal}, with the state $x_{i|k}$ replaced by a data-driven state estimate constructed from input--output data.
\end{problem}

Since the stage cost of this controller acts on the data-driven state estimate instead of the output, the stability proofs of DeePC do not carry over, and the closed-loop properties need to be established anew, which leads to the following problem.

\begin{problem}\label{prob:control}
For the controller of Problem~\ref{prob:state}, find conditions on the offline data and the controller parameters under which the closed loop with~\eqref{eq:plant} is recursively feasible, satisfies the constraints, and is stable.
\end{problem}

Here, stability refers to convergence of the data-driven state estimate to the origin in the noise-free case and to a neighborhood of the origin under Assumption~\ref{ass:bounded-noise}, with the neighborhood size scaling with the noise bound.

\section{DeoPC Design}\label{sec:deopc}

In this section, we present the DeoPC framework. We first construct a data-driven estimate of the state from past input--output data in Section~\ref{sec:state}. Based on this state, we then formulate the DeoPC optimization problem in Section~\ref{sec:deopc-problem}. Finally, we design the terminal cost in Section~\ref{sec:terminal}.

\subsection{Data-Driven State Estimate Construction}\label{sec:state}

The data-driven state estimate is constructed using the subspace identification method~\cite{vanoverschee1996subspace,verheijen2023handbook}.
Let $X^{\rm d}=[x^{\rm d}_{\Tini}\ \cdots\ x^{\rm d}_{\Td-N}]$ denote the states of the offline trajectory at the first future step of each column of the Hankel matrices.
Stacking~\eqref{eq:plant} over the $N$ future steps of every column gives the following equation
\begin{equation}\label{eq:subspace-eq}
\Yf=\mathcal O_NX^{\rm d}+\mathcal T_N\Uf+N_{\rm f},
\end{equation}
where $\mathcal T_N$ is the block Toeplitz matrix of the Markov parameters $D,CB,CAB,\ldots$ of the plant (see \cite{vanoverschee1996subspace}), and $N_{\rm f}$ collects the effect of the noise $w$ and $v$ in the future block on the outputs.
Multiplying~\eqref{eq:subspace-eq} from the right with the orthogonal projector $\Pi^{\perp}:=I-\Uf^\dagger\Uf$ onto $\ker\Uf$ removes the term in $\Uf$ and gives
\begin{equation}\label{eq:proj}
\Yf\Pi^{\perp}=\mathcal O_NX^{\rm d}\Pi^{\perp} + N_{\rm f}\Pi^{\perp}.
\end{equation}

For better illustration and intuition, we derive the construction for noise-free data, for which $N_{\rm f}=0$, and the same procedure applies in the noisy setting as well.
Since $\Tini\geq\ell$, the rows of $X^{\rm d}$ lie in the row space of the past data $\Wp$~\cite{vanoverschee1996subspace}, so that there exists a matrix $L$ with $\mathcal O_NX^{\rm d}=L\Wp$. Then~\eqref{eq:proj} becomes $\Yf\Pi^\perp=L\Wp\Pi^\perp$.
We solve this equation for $L$ by the pseudoinverse of $\Wp\Pi^\perp$, which gives
\begin{equation}\label{eq:Lw}
L_w:=\Yf\Pi^{\perp}\left(\Wp\Pi^{\perp}\right)_{r_w}^\dagger
=L\,\Wp\Pi^{\perp}\left(\Wp\Pi^{\perp}\right)_{r_w}^\dagger,
\end{equation}
where the pseudoinverse is truncated to the rank $r_w=m\Tini+n$ that $\Wp\Pi^\perp$ has for noise-free data.
If $u^{\rm d}$ is persistently exciting of order $\Tini+N+n$, then $\Wp\Pi^{\perp}(\Wp\Pi^{\perp})_{r_w}^\dagger\Wp=\Wp$, and hence $L_w\Wp=L\Wp=\mathcal O_NX^{\rm d}$.
We then compute the rank-$n$ truncated SVD of $L_w$,
\begin{equation}\label{eq:Lw-svd}
L_{w,n}:=U_n\Sigma_nV_n^{\mathsf T}
=\underbrace{U_n\Sigma_n^{1/2}}_{O_z}\;
\underbrace{\Sigma_n^{1/2}V_n^{\mathsf T}}_{\Ez},
\end{equation}
so that $L_{w,n}\Wp=O_z\,(\Ez\Wp)$, where $O_z$ takes the place of $\mathcal O_N$ and $\Ez\Wp$ that of the state sequence $X^{\rm d}$.
Each column of $\Ez\Wp$ is $\Ez$ times one past window of the offline data, so the data-driven state estimate of the online window $\xi_k$ is computed by
\begin{equation}\label{eq:subspace-state}
\xh_k=\Ez \xi_k.
\end{equation}

The data-driven state estimate has the following properties in the noise-free case:
\begin{lemma}\label{lem:similarity}
Let $w=v=0$, and let $u^{\rm d}$ be persistently exciting of order $\Tini+N+n$.
Then there exists a nonsingular $T\in\R^{n\times n}$, fixed over $k$, such that $\Ez\Wp=TX^{\rm d}$ and
\begin{equation}
\xh_k = T x_k
\label{eq:similarity-state}
\end{equation}
for every past window $\xi_k$ of~\eqref{eq:plant}.
\end{lemma}
\begin{proof}
Let $\calW$ denote the subspace of all past windows of~\eqref{eq:plant}; it has dimension $m\Tini+n=r_w$.
Since $\Tini\geq\ell$, a window determines the state at its end, so there is a matrix $F_p$, set to zero on $\calW^\perp$, with $x_k=F_p\xi_k$ for every window.
In particular, $X^{\rm d}=F_p\Wp$, and~\eqref{eq:subspace-eq} reads $\Yf=\mathcal O_NF_p\Wp+\mathcal T_N\Uf$.
By the fundamental lemma, every window in $\calW$, continued with a zero future input, equals $\Wp g$ for some $g\in\ker\Uf$, so the columns of $\Wp\Pi^\perp$ span $\calW$ and $\Wp\Pi^\perp(\Wp\Pi^\perp)_{r_w}^\dagger$ is the orthogonal projector onto $\calW$.
With $\Uf\Pi^\perp=0$, this gives $L_w=\mathcal O_NF_p$.
Both factors have rank $n$, so $L_{w,n}=L_w$, and the two rank-$n$ factorizations $O_z\Ez=\mathcal O_NF_p$ give $\Ez=TF_p$ for a nonsingular $T$.
Hence $\Ez\Wp=TX^{\rm d}$ and $\Ez\xi_k=Tx_k$.
\end{proof}

\subsection{DeoPC Optimization Problem}\label{sec:deopc-problem}

Based on the data-driven state estimate, we now introduce the DeoPC optimization problem.
For $i=0,\dots,N$, let
\begin{equation}\label{eq:Wi}
W_i=\begin{bmatrix}[H_L(u^{\rm d})]_{i:i+\Tini-1}\\ [H_L(y^{\rm d})]_{i:i+\Tini-1}\end{bmatrix},
\end{equation}
where $[\cdot]_{i:j}$ denotes the block rows $i,\dots,j$ of a matrix, with the block rows numbered from $0$, so that $W_0=\Wp$.
The data state matrices are $\Xh_i=\Ez W_i$, $i=0,\dots,N$, and we set $\Xh_{\rm f}=\col(\Xh_1,\dots,\Xh_N)$ and $\calA=\col(\Xh_0,\Uf)$.

At time $k$, given the window $\xi_k$ and $\xh_k=\Ez\xi_k$, let $u=\col(u_{0|k},\dots,u_{N-1|k})$, $\hat{\mathbf x}=\col(\xh_{1|k},\dots,\xh_{N|k})$, and $y=\col(y_{0|k},\dots,y_{N-1|k})$ denote the predicted inputs, data-driven state estimates, and outputs.
The DeoPC problem is formulated as follows:
\begin{equation}\label{eq:qp}
\begin{aligned}
V(\xh_k)=\min_{g,\zeta,u,\hat{\mathbf x},y}\ &\sum_{i=0}^{N-1}\Big(\wnrm{\xh_{i|k}}{Q}+\wnrm{u_{i|k}}{R}\Big)+\lambda\wnrm{\xh_{N|k}}{P}\\
\text{s.t. }&\begin{bmatrix}\Xh_0\\ \Uf\\ \Xh_{\rm f}\\ \Yf\end{bmatrix}g=\begin{bmatrix}\xh_k\\ u\\ \hat{\mathbf x}\\ y\end{bmatrix},\\
&g=\calA^\T\zeta,\quad \xh_{0|k}=\xh_k,\\
&u_{i|k}\in\U,\quad y_{i|k}\in\Ys,\quad i=0,\ldots,N-1,
\end{aligned}
\end{equation}
where $Q\succeq0$ and $R\succ0$ are the stage weights on the data-driven state estimate and the input, $P\succ0$ is the terminal weight with multiplier $\lambda>0$, $\zeta\in\R^{n+mN}$, and 
\begin{equation}\label{eq:robust-margin}
\Ys=\{y:y+s\B\subseteq\Y\}
\end{equation}
is the output constraint set tightened by a margin $s\geq0$.
The input $u_k=u^*_{0|k}$ is applied to the plant, and $V(\xh_k)$ denotes the optimal value, with $V(\xh_k)=+\infty$ when~\eqref{eq:qp} is infeasible.

\begin{remark}\label{rem:g}
As in DeePC~\eqref{eq:deepc}, $g$ combines the columns of the data matrices, and the first two block rows of the equality in~\eqref{eq:qp}, i.e., $\calA g=\col(\xh_k,u)$, fix the data-driven state estimate and the future inputs.
This equation has many solutions $g$, and the constraint $g=\calA^\T\zeta$ selects the one of least norm: when $\calA$ has full row rank, $g$ satisfies both constraints if and only if it solves $\min_g\ \tfrac12\nrm{g}^2 \ \text{s.t. }\calA g=\col(\xh_k,u)$, 
that is, $g=\calA^\dagger\col(\xh_k,u)$, and $\zeta$ is the Lagrange multiplier of this problem.
For noise-free data and under the excitation condition of Lemma~\ref{lem:similarity}, all solutions give the same predicted states and outputs, so the constraint changes nothing.
For noisy data, the solutions can give different predictions, and the constraint makes the prediction a unique linear function of $(\xh_k,u)$.
\end{remark}

\subsection{Terminal Cost and Certificate}\label{sec:terminal}

In the DeoPC problem~\eqref{eq:qp}, the terminal weight $P$ is a design parameter and the closed-loop stability depends on its choice, as is the case in model-based MPC~\cite{rawlings2017model}. 
Since~\eqref{eq:qp} has no terminal constraint, the terminal cost $\lambda\wnrm{\xh_{N|k}}{P}$ has to bound the cost that a local control law $u=K_c\xh$ with a gain $K_c$ incurs after the end of the horizon.
To state this requirement, we first derive the one-step evolution of the data-driven state estimate from the prediction in~\eqref{eq:qp}.
By Remark~\ref{rem:g}, the first predicted state is $\xh_{1|k}=\Xh_1\calA^\dagger\col(\xh_k,u)$.
The first $n+m$ columns of $\Xh_1\calA^\dagger$ form the one-step predictor
\begin{equation}\label{eq:M}
M=[M_x\ M_u]:=\Xh_1\calA^\dagger\begin{bmatrix}I_{n+m}\\0\end{bmatrix},
\end{equation}
so that $M\col(\xh,u_0)$ predicts the next data-driven state.

We write the stage cost as $\ell_c(\xh,u)=\wnrm{\xh}{Q}+\wnrm{u}{R}=\nrm{C_c\col(\xh,u)}^2$ with $C_c=\diag(Q^{1/2},R^{1/2})$.
The local law $u=K_c\xh$ applied to $M$ gives the closed-loop matrix $F=M_x+M_uK_c$ and the stage cost $\ell_c(\xh,K_c\xh)=\nrm{C_K\xh}^2$ with $C_K=C_c\col(I,K_c)$.
For chosen constants $\mu>0$, $\bar p>0$, and $\bar k>0$, the \emph{terminal condition} on $(P,K_c)$ is
\begin{subequations}\label{eq:term}
\begin{align}
&\lambda(F^\T PF-P)+C_K^\T C_K\preceq-\mu I,\label{eq:cert-terminal}\\
&0\prec P\preceq\bar pI,\quad \nrm{K_c}\leq\bar k.\label{eq:term-bounds}
\end{align}
\end{subequations}
Inequality~\eqref{eq:cert-terminal} requires that the terminal cost decreases by at least the stage cost in every step under $u=K_c\xh$.

The pair $(P,K_c)$ is computed as follows.
With $S_P=P^{-1}$, $Y_P=K_cS_P$, $Z_P=M_xS_P+M_uY_P$, and $\mathcal D_P=\col(C_c\col(S_P,Y_P)/\sqrt\lambda,\sqrt{\mu/\lambda}\,S_P)$, a congruence by $S_P$ and a Schur complement show that~\eqref{eq:cert-terminal} together with $0\prec P\preceq\bar pI$ is equivalent to the linear matrix inequality (LMI)
\begin{equation}\label{eq:C-lmi}
\begin{bmatrix}S_P&Z_P^\T&\mathcal D_P^\T\\ Z_P&S_P&0\\ \mathcal D_P&0&I\end{bmatrix}\succeq0,\quad S_P\succeq\bar p^{-1}I,
\end{equation}
whose solution gives $P=S_P^{-1}$ and $K_c=Y_PS_P^{-1}$.
The gain bound $\nrm{K_c}\leq\bar k$ is not convex in $(S_P,Y_P)$, and we replace it by the sufficient LMI
\begin{equation}\label{eq:cert-gain}
\begin{bmatrix}(\bar k/\bar p)I&Y_P\\Y_P^\T&(\bar k/\bar p)I\end{bmatrix}\succeq0,
\end{equation}
that is, $\nrm{Y_P}\leq\bar k/\bar p$, which,  with $\nrm{S_P^{-1}}\leq\bar p$, gives $\nrm{K_c}\leq\bar k$.
Thus every solution of~\eqref{eq:C-lmi} and~\eqref{eq:cert-gain} gives a pair $(P,K_c)$ that satisfies~\eqref{eq:term}.

Since $Q$ may be only positive semidefinite, the stage cost need not bound the data-driven state.
We therefore require in addition that every trajectory of $M$ with zero stage cost converges to the origin, and we state this requirement through a storage function $\xh^\T H_s\xh$.
With $E_x=[I\ 0]$ and for chosen constants $\varepsilon_0>0$ and $0<h_-\leq h_+$, the \emph{storage condition} on $(H_s,\varepsilon_s)$ is
\begin{subequations}\label{eq:cert}
\begin{align}
&M^\T H_sM-E_x^\T H_sE_x+\varepsilon_sE_x^\T E_x\preceq\tfrac12C_c^\T C_c,\label{eq:cert-storage}\\
&h_-I\preceq H_s\preceq h_+I,\quad \varepsilon_s\geq\varepsilon_0.\label{eq:cert-bounds}
\end{align}
\end{subequations}
The storage condition is an LMI in $(H_s,\varepsilon_s)$.
We call the terminal condition~\eqref{eq:term} and the storage condition~\eqref{eq:cert} together the \emph{certificate} on $(H_s,\varepsilon_s,P,K_c)$.
The complete DeoPC scheme is summarized in Algorithm~\ref{alg:deopc}.

\begin{algorithm}
\caption{DeoPC}\label{alg:deopc}
\begin{algorithmic}[1]
\Require offline data $(u^{\rm d},y^{\rm d})$; order $n$; horizons $\Tini$ and $N$; sets $\U$ and $\Y$.
\Statex \textbf{Parameters:} weight $R\succ0$; multiplier $\lambda>0$; margin $s\geq0$; constants $\mu$, $\bar p$, and $\bar k$ of~\eqref{eq:term} and $\varepsilon_0$, $h_-$, and $h_+$ of~\eqref{eq:cert}.
\Statex \textbf{Offline Phase}
\State Build $\Wp$, $\Uf$, $\Yf$ by~\eqref{eq:partition} and $W_i$, $i=0,\ldots,N$, by~\eqref{eq:Wi}.\label{alg:hankel}
\State Compute $\Ez$ by~\eqref{eq:Lw} and~\eqref{eq:Lw-svd}, $\Xh_i=\Ez W_i$, and $\calA$; if $\calA$ lacks full row rank, collect other offline data; otherwise compute $M$ by~\eqref{eq:M}.\label{alg:pred}
\State Choose a weight $Q\succeq0$ on the data-driven state estimate $\xh=\Ez\xi$.\label{alg:weight}
\State Solve~\eqref{eq:C-lmi} and~\eqref{eq:cert-gain} for $(P,K_c)$ and~\eqref{eq:cert} for $(H_s,\varepsilon_s)$.\label{alg:cert}
\Statex \textbf{Online Phase}
\For{$k=0,1,2,\ldots$}\label{alg:online}
\State Measure $\xi_k$, compute $\xh_k=\Ez\xi_k$, solve~\eqref{eq:qp}, and apply $u_k=u^*_{0|k}$.\label{alg:apply}
\EndFor
\end{algorithmic}
\end{algorithm}

\begin{remark}\label{rem:deepc}
For system~\eqref{eq:plant} with $D=0$, DeePC~\eqref{eq:deepc} can be viewed as a special case of the proposed DeoPC~\eqref{eq:qp} with $Q=T^{-\T}C^\T SCT^{-1}$ and without terminal cost ($\lambda=0$).
Indeed, under the conditions of Lemma~\ref{lem:similarity}, we have $y_k=CT^{-1}\xh_k$ along every trajectory of~\eqref{eq:plant}, so that $\wnrm{y_k}{S}=\wnrm{\xh_k}{Q}$ and the two costs coincide.
This weight has rank at most $p$ and acts only on the output, whereas $Q$ in~\eqref{eq:qp} is an arbitrary positive semidefinite matrix.
\end{remark}

\section{Theoretical Properties}\label{sec:main}

In this section, we analyze the theoretical properties of the proposed DeoPC framework.
We begin with the noise-free case in Section~\ref{sec:nominal}, where we compare DeoPC with state-feedback MPC and analyze the stability of the closed loop.
We then turn to the more general case of noisy data in Section~\ref{sec:noisy} and analyze the closed-loop stability under bounded noise.
To facilitate the theoretical analysis, we make the following assumption, which gives the constraint sets an explicit representation.

\begin{assumption}\label{ass:constraints}
The constraint sets can be described in the form of polytopes
\begin{equation}
    \U = \{u:H_u u \leq h_u\},\quad
    \Y = \{y:H_y y \leq h_y\},
\end{equation}
where the rows of $H_u$ and $H_y$ are nonzero.
\end{assumption}

Under Assumption~\ref{ass:constraints}, the largest balls centered at the origin that $\U$ and $\Y$ contain have the radii $r_\U=\min_j(h_u)_j/\nrm{(H_u)_j}$ and $r_\Y=\min_j(h_y)_j/\nrm{(H_y)_j}$.

\subsection{Noise-Free Data}\label{sec:nominal}

When noise-free data are considered, i.e., $w=v=0$, we have $\xh_k=Tx_k$ by Lemma~\ref{lem:similarity}.
The output constraint set~\eqref{eq:robust-margin} need not be tightened, and we set the margin to $s=0$.
We denote by $A_z=TAT^{-1}$, $B_z=TB$, and $C_z=CT^{-1}$ the plant matrices in the coordinates $\xh=Tx$.
The following theorem shows that, in this case, DeoPC is equivalent to the model-based state-feedback MPC~\eqref{eq:ideal} reviewed in Section~\ref{sec:mpc}.

\begin{theorem}\label{prop:equiv}
Consider system~\eqref{eq:plant} in the noise-free case, i.e., $w=v=0$.
Suppose that $u^{\rm d}$ is persistently exciting of order $\Tini+N+n$.
Then, for every past window $\xi_k$ of~\eqref{eq:plant}, DeoPC~\eqref{eq:qp} is equivalent to the state-feedback MPC~\eqref{eq:ideal} with $Q_x=T^\T QT$ and $P_x=\lambda T^\T PT$, in the sense that the two problems have the same feasible input sequences, the same optimal value, and the same optimal input sequence.
\end{theorem}
\begin{proof}
For noise-free data, the columns of $\Xh_i$ are $T$ times states of the offline trajectory, so $\Xh_{i+1}=A_z\Xh_i+B_z U_i$ and $Y_i=C_z\Xh_i+DU_i$, where $U_i$ and $Y_i$ denote block row $i$ of $\Uf$ and $\Yf$.
Hence, for every $d=\col(\xh,u_0,\ldots,u_{N-1})$, every $g$ with $\calA g=d$ satisfies
\begin{equation}\label{eq:exactpred}
\begin{aligned}
\Xh_ig&=A_z^i\xh+\sum_{j<i}A_z^{i-1-j}B_zu_j,\quad i=0,\ldots,N,\\
Y_ig&=C_z\Xh_ig+Du_i,\quad i=0,\ldots,N-1,
\end{aligned}
\end{equation}
and such a $g$ exists for every $d$, since $\calA$ has full row rank \cite{willems2005note}.
Every $g$ that satisfies the equality constraints of~\eqref{eq:qp} satisfies $\calA g=\col(\xh_k,u)$, so, by~\eqref{eq:exactpred} and $\xh_k=Tx_k$, $\xh_{i|k}=\Xh_ig$ and $y_{i|k}=Y_ig$ equal $Tx_{i|k}$ and $Cx_{i|k}+Du_{i|k}$ with the states $x_{i|k}$ of~\eqref{eq:ideal} under the same inputs.
Conversely, since $\calA$ has full row rank, $g=\calA^\dagger\col(\xh_k,u)\in\operatorname{im}\calA^\T$ satisfies these constraints for every $u$.
Since $\Ys=\Y$ for $s=0$, $\wnrm{Tx}{Q}=\wnrm{x}{T^\T QT}$, and $\lambda\wnrm{Tx}{P}=\wnrm{x}{\lambda T^\T PT}$, both problems have the same constraints and the same cost on every input sequence, which gives the claims.
\end{proof}

Theorem~\ref{prop:equiv} shows that, for noise-free data, the weight $Q$ on the data-driven state estimate acts as the weight $T^\T QT$ on the internal state, and that DeoPC inherits the optimal inputs of the corresponding state-feedback MPC.
With this property, we now turn to the stability of the closed loop, which can be analyzed along the lines of model-based MPC \cite{rawlings2017model} by means of the certificate of Section~\ref{sec:terminal}.
We first characterize the feasibility of the certificate in the following lemma.

\begin{lemma}\label{lem:storage}
Consider system~\eqref{eq:plant} in the noise-free case, and suppose that $u^{\rm d}$ is persistently exciting of order $\Tini+N+n$.
Then the certificate given by~\eqref{eq:term} and~\eqref{eq:cert} is feasible for some constants $\mu,\varepsilon_0,h_-,h_+,\bar p,\bar k$ if and only if $(Q^{1/2},A_z)$ is detectable.
\end{lemma}
\begin{proof}
For noise-free data, \eqref{eq:exactpred} with $i=1$ gives $\Xh_1\calA^\dagger\col(\xh,u)=A_z\xh+B_zu_0$, and hence the one-step predictor~\eqref{eq:M} is $M=[A_z\ B_z]$.
Suppose first that the certificate is feasible for some constants. Then~\eqref{eq:cert-bounds} gives $H_s\succ0$ and $\varepsilon_s>0$, and~\eqref{eq:cert-storage} with $u=0$ gives $A_z^\T H_sA_z-H_s+\varepsilon_sI\preceq\frac12Q$; for an eigenvector $\chi$ with $A_z\chi=\vartheta\chi$ and $Q^{1/2}\chi=0$, this yields $(|\vartheta|^2-1)\chi^*H_s\chi+\varepsilon_s\nrm{\chi}^2\leq0$, so $|\vartheta|<1$, and the PBH test gives detectability.

Conversely, since $(Q^{1/2},A_z)$ is detectable, it admits a quadratic IOSS Lyapunov function \cite{berberich2021data}, that is, there exist $H_0\succ0$ and constants $c_q,c_u>0$ such that
\begin{equation}
\begin{aligned}
&(A_z\xh+B_zu)^\T H_0(A_z\xh+B_zu)-\xh^\T H_0\xh\\
&\qquad\leq-\tfrac12\nrm{\xh}^2+c_q\xh^\T Q\xh+c_u\nrm{u}^2
\end{aligned}    
\end{equation}
holds for all $\xh$ and $u$.
With $H_s=\theta H_0$, $\theta>0$, $\theta c_q<\tfrac12$, $\theta c_u I\prec\tfrac12 R$, and $\varepsilon_s=\theta/4$, \eqref{eq:cert-storage} holds.
Since $(A_z,B_z)$ is controllable, some $K_c$ makes $F=A_z+B_zK_c$ Schur, and for every $\mu>0$ the discrete Lyapunov equation $\lambda(F^\T PF-P)=-C_K^\T C_K-\mu I$ has a solution $P\succ0$, which satisfies~\eqref{eq:cert-terminal}.
The bounds~\eqref{eq:term-bounds} and~\eqref{eq:cert-bounds} then hold for suitable constants.
\end{proof}

Lemma~\ref{lem:storage} shows that the certificate is feasible exactly when every mode of the plant that is not penalized by $Q$ is stable.
In particular, a certificate exists for every $Q\succ0$, without any knowledge of the plant.

We next define the regions of initial states on which the stability result holds.
Let $a_j$ denote the $j$th row of $\col(H_uK_c,H_y(C_z+DK_c))$ and $b_j$ the $j$th entry of $\col(h_u,h_y)$, and let
\begin{equation}\label{eq:B-alpha}
\alpha=\min_{j:\,a_jP^{-1}a_j^\T>0}\frac{b_j^2}{a_jP^{-1}a_j^\T},
\end{equation}
which is the largest level of the ellipsoid $\Omega=\{\xh:\xh^\T P\xh\leq\alpha\}$ on which the local law $u=K_c\xh$ satisfies the input and output constraints.
With $\eta=\lambda_{\min}(P^{-1/2}QP^{-1/2})$ and $\delta=\eta\alpha$, the regions are given by
\begin{equation}\label{eq:regions}
\mathcal X^0_{N,\lambda}=\{\xh:V(\xh)\leq\lambda\alpha\},\quad
\mathcal X_{N,\lambda}=\{\xh:V(\xh)\leq N\delta+\lambda\alpha\},
\end{equation}
and the stability result for noise-free data is stated as follows.

\begin{theorem}\label{thm:nominal}
Consider system~\eqref{eq:plant} in the noise-free case under Assumption~\ref{ass:constraints}.
Suppose that $u^{\rm d}$ is persistently exciting of order $\Tini+N+n$ and that $(H_s,\varepsilon_s,P,K_c)$ satisfies the certificate given by~\eqref{eq:term} and~\eqref{eq:cert}.
Then there exist $c>0$ and $\rho\in(0,1)$ such that, for every initial window $\xi_0$ of~\eqref{eq:plant} with $\xh_0\in\mathcal X^0_{N,\lambda}$, the closed loop of~\eqref{eq:plant} and~\eqref{eq:qp} is recursively feasible, satisfies $u_k\in\U$ and $y_k\in\Y$ for every $k\geq0$, and satisfies
\begin{equation}\label{eq:thm1-bound}
\nrm{\xh_k}\leq c\,\rho^k\nrm{\xh_0},\quad k \geq 0.
\end{equation}
If $Q\succ0$, the same holds for initial $\xh_0\in\mathcal X_{N,\lambda}$.
\end{theorem}
\begin{proof}
By Theorem~\ref{prop:equiv} and~\eqref{eq:exactpred}, problem~\eqref{eq:qp} is the MPC problem for $(A_z,B_z,C_z,D)$ with the state $\xh$, the closed loop obeys $\xh_{k+1}=A_z\xh_k+B_zu_k$ and $y_k=C_z\xh_k+Du_k$, and $M=[A_z\ B_z]$, so that $F=A_z+B_zK_c$.
The proof therefore follows the arguments for MPC without terminal constraint \cite{limon2006stability,rawlings2017model} and proceeds in three steps.

First, we show that $\Omega$ is invariant under the local law $u=K_c\xh$ and contains the terminal state $\xh_N^*$ of the optimal prediction from every $\xh$ in the regions~\eqref{eq:regions}.
The level $\alpha$ in~\eqref{eq:B-alpha} satisfies $0<\alpha<\infty$, since each $b_j$ is positive because the origin is interior to both polytopes, and some denominator is positive because otherwise the boundedness of $\U$ and $\Y$ would give $K_c=0$ and $C_z=0$, which contradicts observability.
By the definition of $\alpha$, the local law satisfies the input and output constraints on $\Omega$.
Moreover, the terminal condition~\eqref{eq:cert-terminal} gives
\begin{equation}\label{eq:terminal}
\lambda\wnrm{F\xh}{P}-\lambda\wnrm{\xh}{P}
\le-\ell_c(\xh,K_c\xh),
\end{equation}
so that $\Omega$ is invariant under the local law.
Applying the local law over the whole horizon from $\xh\in\Omega$ is thus feasible, and summing~\eqref{eq:terminal} along the resulting trajectory yields $V(\xh)\le\lambda\wnrm{\xh}{P}$ for every $\xh\in\Omega$.
If $V(\xh)\le\lambda\alpha$, then $\lambda\wnrm{\xh_N^*}{P}\le V(\xh)\le\lambda\alpha$, and hence $\xh_N^*\in\Omega$.
If $Q\succ0$ and $V(\xh)\le N\delta+\lambda\alpha$, then the optimal prediction enters $\Omega$ at some stage $j\le N$, since $\ell_c(\xh,u)\ge\eta\wnrm{\xh}{P}>\delta$ outside $\Omega$ and its cost would otherwise exceed $N\delta+\lambda\alpha$.
Since the tail of the optimal prediction from stage $j$ is optimal for the horizon $N-j$, whose optimal value is denoted by $V_{N-j}$, and the bound of the local law holds for every horizon, we obtain
\begin{equation*}
\lambda\wnrm{\xh_N^*}{P}\le V_{N-j}(\xh_j^*)\le\lambda\wnrm{\xh_j^*}{P}\le\lambda\alpha,
\end{equation*}
and hence $\xh_N^*\in\Omega$ in this case as well.

Next, we prove recursive feasibility and constraint satisfaction.
Let $\xh_k \in \mathcal X^0_{N,\lambda}$, or $\xh_k \in \mathcal X_{N,\lambda}$ if $Q\succ0$.
Shifting the optimal input sequence at time $k$ by one stage and appending $K_c\xh_N^*$ gives a feasible input sequence at time $k+1$, since its first $N-1$ stages are stages of the optimal prediction at time $k$ and the appended stage satisfies the constraints because $\xh_N^*\in\Omega$.
Comparing the cost of this sequence with $V(\xh_k)$ and using~\eqref{eq:terminal} at $\xh_N^*$ yields
\begin{equation}\label{eq:nomdec}
V(\xh_{k+1})-V(\xh_k)\le-\ell_c(\xh_k,u_k).
\end{equation}
Hence $\mathcal X^0_{N,\lambda}$ and, if $Q\succ0$, $\mathcal X_{N,\lambda}$ are invariant, so that the argument applies at every $k$ and the closed loop is recursively feasible.
Moreover, $u_k\in\U$ and $y_k\in\Y$ hold for every $k\geq0$, since $u_k$ and $y_k=C_z\xh_k+Du_k$ are the input and the output of the first stage of the optimal prediction.

Finally, we prove the exponential bound~\eqref{eq:thm1-bound}.
For $Q\succeq0$ and $\xh_0\in\mathcal X^0_{N,\lambda}$, adding the storage condition~\eqref{eq:cert-storage} at $(\xh_k,u_k)$, where $\xh_{k+1}=M\col(\xh_k,u_k)$, to~\eqref{eq:nomdec} gives, for
\begin{equation}\label{eq:VH}
V_H(\xh)=V(\xh)+\xh^\T H_s\xh,
\end{equation}
the decrease
\begin{equation}\label{eq:nomstore}
V_H(\xh_{k+1})-V_H(\xh_k)
\le-\tfrac12\ell_c(\xh_k,u_k)-\varepsilon_s\nrm{\xh_k}^2.
\end{equation}
On $\mathcal X^0_{N,\lambda}$, the bound $V(\xh)\le\lambda\wnrm{\xh}{P}$ holds inside $\Omega$ by the first step and outside $\Omega$ because $V(\xh)\le\lambda\alpha<\lambda\wnrm{\xh}{P}$ there.
Hence $\lambda_{\min}(H_s)\nrm{\xh}^2\le V_H(\xh)\le c_H\nrm{\xh}^2$ with $c_H=\lambda\lambda_{\max}(P)+\lambda_{\max}(H_s)$, and~\eqref{eq:nomstore} yields $V_H(\xh_{k+1})\leq(1-\varepsilon_s/c_H)V_H(\xh_k)$, which gives~\eqref{eq:thm1-bound} with $\rho=(1-\varepsilon_s/c_H)^{1/2}$ and $c=(c_H/\lambda_{\min}(H_s))^{1/2}$.
For $Q\succ0$ and $\xh_0\in\mathcal X_{N,\lambda}$, the first stage cost gives $V(\xh)\ge\lambda_{\min}(Q)\nrm{\xh}^2$, and $V(\xh)\le c_V\nrm{\xh}^2$ holds on $\mathcal X_{N,\lambda}$ with $c_V=(\lambda+N\eta)\lambda_{\max}(P)$, inside $\Omega$ by the first step and outside $\Omega$ because $V(\xh)\le N\delta+\lambda\alpha<(N\eta+\lambda)\wnrm{\xh}{P}$ there.
Hence~\eqref{eq:nomdec} yields $V(\xh_{k+1})\leq(1-\lambda_{\min}(Q)/c_V)V(\xh_k)$, which gives~\eqref{eq:thm1-bound} with $\rho=(1-\lambda_{\min}(Q)/c_V)^{1/2}$ and $c=(c_V/\lambda_{\min}(Q))^{1/2}$.
\end{proof}

\begin{remark}\label{rem:region}
The condition $\xh_0\in\mathcal X^0_{N,\lambda}$ in Theorem~\ref{thm:nominal} can be checked at $k=0$ from the data alone.
Indeed, $V(\xh_0)$ is the optimal value of~\eqref{eq:qp} at $k=0$, and $\alpha$ in~\eqref{eq:B-alpha} is computed from $(P,K_c)$, the constraint sets, and $[C_z\ D]$, which by~\eqref{eq:exactpred} consists of the columns of $Y_0\calA^\dagger$ that act on $(\xh,u_0)$.
\end{remark}

\subsection{Noisy Data}\label{sec:noisy}

Having analyzed the closed-loop stability for noise-free data, we now turn to the more general case in which the offline and online data are affected by the bounded noise of Assumption~\ref{ass:bounded-noise}.
To describe the effect of the process noise $w$ and the measurement noise $v$ by a single signal, we fix a gain $K$ such that $\Phi=A-KC$ is Schur, which exists since $(C,A)$ is observable, together with constants $c_\Phi\geq1$ and $\rho_\Phi\in(0,1)$ such that $\nrm{\Phi^j}\leq c_\Phi\rho_\Phi^j$ for every $j\geq0$.
For a trajectory of~\eqref{eq:plant} that starts at time $k_0$, let $\bar x_{k_0}=x_{k_0}$ and, for $k\geq k_0$,
\begin{equation}\label{eq:innov}
e_k=y_k-C\bar x_k-Du_k,
\quad \bar x_{k+1}=A\bar x_k+Bu_k+Ke_k,
\end{equation}
so that $y_k=C\bar x_k+Du_k+e_k$, which is the innovation form that is common in subspace identification \cite{vanoverschee1996subspace}.
The following lemma bounds the signal $e_k$ and the deviation of $\bar x_k$ from the state $x_k$.

\begin{lemma}\label{lem:innov}
Let Assumption~\ref{ass:bounded-noise} hold.
Then, for every trajectory of~\eqref{eq:plant} that starts at time $k_0$, the signals in~\eqref{eq:innov} satisfy $\nrm{x_k-\bar x_k}\leq\bar\varepsilon$ and $\nrm{e_k}\leq\bar e$ for every $k\geq k_0$, where
\begin{equation}\label{eq:ebar}
\bar\varepsilon=\frac{c_\Phi(\bar w+\nrm K\bar v)}{1-\rho_\Phi},\quad \bar e=\nrm C\bar\varepsilon+\bar v.
\end{equation}
\end{lemma}
\begin{proof}
Subtracting~\eqref{eq:innov} from~\eqref{eq:plant} gives $e_k=C(x_k-\bar x_k)+v_k$ and $x_{k+1}-\bar x_{k+1}=\Phi(x_k-\bar x_k)+w_k-Kv_k$.
Since $x_{k_0}-\bar x_{k_0}=0$, it follows that $x_k-\bar x_k=\sum_{j=k_0}^{k-1}\Phi^{k-1-j}(w_j-Kv_j)$, and the bounds of Assumption~\ref{ass:bounded-noise} together with $\nrm{\Phi^j}\leq c_\Phi\rho_\Phi^j$ yield $\nrm{x_k-\bar x_k}\leq c_\Phi(\bar w+\nrm K\bar v)\sum_{j\geq0}\rho_\Phi^j=\bar\varepsilon$.
Hence $\nrm{e_k}\leq\nrm C\bar\varepsilon+\bar v=\bar e$.
\end{proof}

We refer to $\bar e$ as the \emph{noise level}, which is proportional to the noise bounds and vanishes for noise-free data, in which case $\bar x_k=x_k$.

Before establishing the stability result, we need the following two lemmas, whose proofs are given in Appendices~\ref{app:perturbation} and~\ref{app:state-error}.
The first lemma bounds how the noise in the offline data perturbs the maps that DeoPC builds from them, and the second lemma bounds the resulting error of the data-driven state estimate.
Let
\begin{equation}\label{eq:pred}
\begin{aligned}
\Th_i^x&=\Xh_i\calA^\dagger,\quad i=0,\dots,N,\\
\Th_i^y&=Y_i\calA^\dagger,\quad i=0,\dots,N-1,
\end{aligned}
\end{equation}
denote the prediction maps of~\eqref{eq:qp}, and let $E_0$, $\Th_i^{x,0}$, and $\Th_i^{y,0}$ denote the state map $\Ez$ and the prediction maps that are obtained in the same way from the noise-free response of~\eqref{eq:plant} to the recorded input $u^{\rm d}$ from the same initial state, with a fixed choice of the SVD in~\eqref{eq:Lw-svd}.
For an orthogonal matrix $\Lambda\in\R^{n\times n}$, the error of the noisy maps is measured by
\begin{equation}\label{eq:pert}
\begin{aligned}
t(\Lambda)={}&\nrm{\Lambda^\T\Ez-E_0}+\max_i\nrm{\Lambda^\T\Th_i^x\Delta_\Lambda-\Th_i^{x,0}}\\
&+\max_i\nrm{\Th_i^y\Delta_\Lambda-\Th_i^{y,0}},
\end{aligned}
\end{equation}
where $\Delta_\Lambda=\diag(\Lambda,I_{mN})$.
Moreover, let $\sigma_d=\smin\big(H_{\Tini+N+n}(u^{\rm d})\big)$ denote the excitation level and $\dd=\max_{0\leq k<\Td}\nrm{\col(u^{\rm d}_k,y^{\rm d}_k)}$ the size of the offline data.

\begin{lemma}\label{lem:pert}
Let Assumption~\ref{ass:bounded-noise} hold.
Then there exist constants $C_1,C_2,C_E>0$ and a threshold $\sigma_\star(\ee,\dd;\tau)$ for every $\tau>0$ such that, if $\sigma_d>\sigma_\star(\ee,\dd;\tau)$, then the truncations in~\eqref{eq:Lw} and~\eqref{eq:Lw-svd} are unique, $\calA$ has full row rank, and, for every choice of the SVD in~\eqref{eq:Lw-svd}, there exists an orthogonal matrix $\Lambda$ such that
\begin{equation*}
\nrm{\Ez-\Lambda E_0}\le C_E\frac{\ee}{\sigma_d},\quad
t(\Lambda)\le C_1\frac{\ee}{\sigma_d}+C_2\frac{\ee\dd}{\sigma_d^2}<\tau.
\end{equation*}
Moreover, the constants are independent of the noise realization and of the choice of the SVD, and the threshold satisfies $\sigma_\star(r\ee,r\dd;\tau)=r\,\sigma_\star(\ee,\dd;\tau)$ for every $r>0$ and $\sigma_\star(0,\dd;\tau)=0$.
\end{lemma}

\begin{lemma}\label{lem:state-error}
Let Assumption~\ref{ass:bounded-noise} hold and suppose that $\sigma_d>\sigma_\star(\ee,\dd;\tau)$ for some $\tau>0$. Let $\Lambda$ denote the orthogonal matrix in Lemma~\ref{lem:pert}.
Then there exists a constant $c_x>0$ such that
\begin{equation}\label{eq:state-error}
\nrm{\Lambda^\T\xh_k-Tx_k}\le C_E\frac{\ee}{\sigma_d}\nrm{\xi_k}+c_x\ee
\end{equation}
holds for every past window $\xi_k$ of~\eqref{eq:plant}. In addition, $c_x$ is independent of the noise realization.
\end{lemma}

We are now ready to establish the stability of the closed loop for noisy data. Since the error of $\xh_k$ in~\eqref{eq:state-error} grows with $\nrm{\xi_k}$, we use a Lyapunov function that also weights the past window.
For $\beta\in(0,1]$, let
\begin{equation}\label{eq:W}
\begin{aligned}
\Lyap_k&=V(\xh_k)+\xh_k^\T H_s\xh_k+\beta\varphi_k,\\
\varphi_k&=\sum_{j=1}^{\Tini}(\Tini-j+1)\nrm{\col(u_{k-j},y_{k-j})}^2.
\end{aligned}
\end{equation}
The stability result for noisy data is stated as follows.

\begin{theorem}\label{thm:noisy}
Consider system~\eqref{eq:plant} under Assumptions~\ref{ass:bounded-noise} and~\ref{ass:constraints}, and let the constants in~\eqref{eq:term} and~\eqref{eq:cert} be fixed.
Then there exist a tolerance $t_*\in(0,1]$, constants $a,b,\gamma_0,\gamma>0$, $\beta\in(0,1]$, and $\rho\in(0,1)$, and a level $c(s)>0$ for each $s\in(0,r_\Y)$ such that the following holds.
Suppose that
\begin{enumerate}
\item\label{it:noisy-cert} $(H_s,\varepsilon_s,P,K_c)$ satisfies~\eqref{eq:term} and~\eqref{eq:cert},
\item\label{it:noisy-pe} $u^{\rm d}$ is $\sigma$-persistently exciting of order $\Tini+N+n$ with $\sigma>\sigma_\star(\bar e,\bar d;t_*)$, where $\sigma_\star$ is the threshold of Lemma~\ref{lem:pert},
\item\label{it:noisy-margin} $a\bar e<s<r_\Y-b\bar e$ and $r_\U\geq b\bar e$,
\item\label{it:noisy-init} the initial window $\xi_0$ is generated by~\eqref{eq:plant} and $\Lyap_0\leq c(s)$ for~\eqref{eq:W} with this $\beta$.
\end{enumerate}
Then the closed loop of~\eqref{eq:plant} and~\eqref{eq:qp} is recursively feasible, satisfies $u_k\in\U$ and $y_k\in\Y$ for every $k\geq0$, and satisfies
\begin{equation}\label{eq:iss}
\nrm{\xh_k}\leq\gamma_0\rho^{k/2}\sqrt{\Lyap_0}+\gamma\bar e,\quad k\geq0.
\end{equation}
\end{theorem}
\begin{proof}
We establish the result in three parts.

\proofhead{Part I}
We first bound the optimal solution of~\eqref{eq:qp} and the errors of its one-step predictions in terms of the error of the maps in Lemma~\ref{lem:pert}.
To this end, let $t_*\in(0,1]$ be a tolerance that is fixed in Part~II.
Since condition~\ref{it:noisy-pe}) gives $\sigma_d\geq\sigma>\sigma_\star(\ee,\dd;t_*)$, Lemma~\ref{lem:pert} with $\tau=t_*$ shows that $\calA$ has full row rank and that there exists an orthogonal matrix $\Lambda$ with $t(\Lambda)<t_*$.
We change the coordinates of the data-driven state estimate to $\Lambda^\T\xh$, which replaces $\Ez$, $\Th_i^x$, and $\Th_i^y$ by $\Lambda^\T\Ez$, $\Lambda^\T\Th_i^x\Delta_\Lambda$, and $\Th_i^y\Delta_\Lambda$, the weights $Q$, $P$, and $H_s$ by $\Lambda^\T Q\Lambda$, $\Lambda^\T P\Lambda$, and $\Lambda^\T H_s\Lambda$, and the gain $K_c$ by $K_c\Lambda$.
Since $\Lambda$ is orthogonal, problem~\eqref{eq:qp} has the same optimal inputs and the same optimal value in the new coordinates, the inequalities~\eqref{eq:cert-terminal} and~\eqref{eq:cert-storage} are transformed by congruence and thus remain valid, and the bounds~\eqref{eq:term-bounds} and~\eqref{eq:cert-bounds} as well as all Euclidean norms are unchanged.
Hence we assume $\Lambda=I$ without loss of generality, so that $t:=t(I)<t_*$ and~\eqref{eq:pert} gives, for every $i$,
\begin{equation}\label{eq:terr}
\nrm{\Ez-E_0},\ \nrm{\Th_i^x-\Th_i^{x,0}},\ \nrm{\Th_i^y-\Th_i^{y,0}}\le t.
\end{equation}

We first bound the optimal value from above.
Let $d=\col(\xh,u)$ with $u=\col(u_0,\ldots,u_{N-1})$, and let $J_z$, $J_i$, and $J_u$ denote the matrices that select $\xh$, $u_i$, and $u$ from $d$, so that $\Th_0^x=J_z$.
By~\eqref{eq:pred}, the cost of~\eqref{eq:qp} is $d^\T\J d$ with
$\J=\sum_{i=0}^{N-1}(\Th_i^{x\T}Q\Th_i^x+J_i^\T RJ_i)+\lambda\Th_N^{x\T}P\Th_N^x$,
so that $V(\xh)$ is the minimum of $d^\T\J d$ over the feasible vectors $d$ with $J_zd=\xh$, where $d$ is called feasible if $J_id\in\U$ and $\Th_i^yd\in\Ys$ for $i=0,\ldots,N-1$.
Consider the vector $\Ls\xh$ that results from applying the local law $u_i=K_c\Th_i^xd$ at every stage.
With $K_N=I_N\otimes K_c$, $\Th_{<N}^x=\col(\Th_0^x,\ldots,\Th_{N-1}^x)$, and $G=K_N\Th_{<N}^xJ_u^\T$, the inputs of this vector solve $u=Gu+K_N\Th_{<N}^xJ_z^\T\xh$, so that, if $I-G$ is invertible,
\[
\Ls=J_z^\T+J_u^\T(I-G)^{-1}K_N\Th_{<N}^xJ_z^\T
\]
and $J_z\Ls=I$.
For the noise-free maps, \eqref{eq:exactpred} shows that the state at stage $i$ does not depend on $u_j$ for $j\ge i$, so that $G_0=K_N\Th_{<N}^{x,0}J_u^\T$ is strictly block lower triangular and $I-G_0$ is invertible.
Since $\nrm{G-G_0}\le\sqrt N\bar kt$ by~\eqref{eq:terr}, there exist constants $t_1\in(0,1]$ and $L_b\geq1$ such that $I-G$ is invertible and $\nrm\Ls\le L_b$ whenever $t\le t_1$, and we require $t_*\leq t_1$.

The vector $\Ls\xh$ is feasible when $\xh$ is small.
Indeed, let $o_b\geq1$ be a constant such that $\nrm{\Th_i^y}\leq o_b$ for every $i$, which exists by~\eqref{eq:terr} with $t\leq1$, and let
\begin{equation}\label{eq:delta}
\bar\delta(s)=\min\{r_\U,(r_\Y-s)/o_b\},\quad
\varrho_0(s)=\bar\delta(s)/(2L_b).
\end{equation}
Every $d$ with $\nrm d\le\bar\delta(s)$ is feasible, since $\nrm{J_id}\le r_\U$ gives $J_id\in\U$ and $\nrm{\Th_i^yd}\le r_\Y-s$ gives $\Th_i^yd+s\B\subseteq r_\Y\B\subseteq\Y$.
Hence, for $\nrm\xh\le\varrho_0(s)$, the vector $\Ls\xh$ is feasible because $\nrm{\Ls\xh}\le\bar\delta(s)/2$, and $V_H(\xh)\le c_+\nrm\xh^2$ holds for $V_H$ in~\eqref{eq:VH} with $c_+=j_bL_b^2+h_+$, where $j_b$ is a constant such that $\nrm\J\leq j_b$.
Moreover, the optimizer $d^*(\xh)$ of~\eqref{eq:qp} is unique since $R\succ0$, and $\J\succeq J_u^\T(I_N\otimes R)J_u$ together with $H_s\succeq h_-I$ gives $V_H(\xh)\ge h_0\nrm{d^*(\xh)}^2$ with $h_0=\min\{h_-,\lambda_{\min}(R)\}$.
Hence, with $\nu=\sqrt{c_+/h_0}$,
\begin{equation}\label{eq:local}
h_0\nrm{d^*(\xh)}^2\le V_H(\xh)\le c_+\nrm{\xh}^2,
\quad\nrm{d^*(\xh)}\le\nu\nrm{\xh}.
\end{equation}

Next, let $\nrm{\xh_k}\le\varrho_0(s)$ and $d_k^*=d^*(\xh_k)$, and consider the errors of the predictions $\Th_1^xd_k^*$ and $\Th_0^yd_k^*$ of the next state and of the current output.
We use the innovation form~\eqref{eq:innov} with $k_0=-\Tini$, for which $\nrm{e_k}\le\ee$ by Lemma~\ref{lem:innov}, and let $\mathbf e_k=\col(e_{k-\Tini},\ldots,e_{k-1})$, so that $\nrm{\mathbf e_k}\le\sqrt{\Tini}\,\ee$.
By the identity~\eqref{eq:innovation} in the proof of Lemma~\ref{lem:state-error}, the error $\tilde x_k=\xh_k-T\bar x_k$ of the data-driven state estimate satisfies $\tilde x_k=(\Ez-E_0)\xi_k+R_v\mathbf e_k$ with a fixed matrix $R_v$, and hence $\nrm{\tilde x_k}\le t\nrm{\xi_k}+\nrm{R_v}\nrm{\mathbf e_k}$ by~\eqref{eq:terr}.
Multiplying~\eqref{eq:innov} by $T$ gives $T\bar x_{k+1}=A_zT\bar x_k+B_zu_k+TKe_k$ and $y_k=C_zT\bar x_k+Du_k+e_k$, and~\eqref{eq:exactpred} gives $\Th_1^{x,0}d_k^*=A_z\xh_k+B_zu_k$ and $\Th_0^{y,0}d_k^*=C_z\xh_k+Du_k$.
Substituting $T\bar x_k=\xh_k-\tilde x_k$ yields $y_k=C_z(\xh_k-\tilde x_k)+Du_k+e_k$ and, for the mismatch $\omega_k=\xh_{k+1}-\Th_1^xd_k^*$ between the measured and the predicted next state,
\begin{align*}
\omega_k&=(\Th_1^{x,0}-\Th_1^x)d_k^*+\tilde x_{k+1}-A_z\tilde x_k+TKe_k,\\
y_k-\Th_0^yd_k^*&=(\Th_0^{y,0}-\Th_0^y)d_k^*-C_z\tilde x_k+e_k.
\end{align*}
Since $\nrm{u_k}\le\nrm{d_k^*}\le\nu\nrm{\xh_k}$ by~\eqref{eq:local}, the expression for $y_k$ gives, for some constants $c_{yz}$, $c_{yw}$, and $c_{yv}$,
\begin{equation}\label{eq:ybound}
\nrm{y_k}\le c_{yz}\nrm{\xh_k}+c_{yw}t\nrm{\xi_k}+c_{yv}\nrm{\mathbf e_k}+\nrm{e_k}.
\end{equation}
Moreover, since $\xi_{k+1}$ and $\mathbf e_{k+1}$ are obtained from $\xi_k$ and $\mathbf e_k$ by dropping the oldest entries and appending $(u_k,y_k)$ and $e_k$, the bound on $\tilde x_k$ gives $\nrm{\tilde x_{k+1}}\le t(\nrm{\xi_k}+\nrm{u_k}+\nrm{y_k})+\nrm{R_v}(\nrm{\mathbf e_k}+\nrm{e_k})$.
Inserting these bounds and~\eqref{eq:terr}, \eqref{eq:local}, and $t\leq1$ into the two expressions above, we have the following bounds hold for some constants $a_{\omega z}$, $a_{\omega w}$, $E_v$, $E_e$, $a_{yz}$, $a_{yw}$, $B_v$, and $B_e$:
\begin{align}\label{eq:noise}
\nrm{\omega_k}&\le t(a_{\omega z}\nrm{\xh_k}+a_{\omega w}\nrm{\xi_k})
    \nonumber\\
&\quad+E_v\nrm{\mathbf e_k}+E_e\nrm{e_k},\nonumber\\
\nrm{y_k-\Th_0^yd_k^*}
&\le t(a_{yz}\nrm{\xh_k}+a_{yw}\nrm{\xi_k})
    \nonumber\\
&\quad+B_v\nrm{\mathbf e_k}+B_e\nrm{e_k}.
\end{align}

\proofhead{Part II}
We then show that the decrease~\eqref{eq:decrease} of $\Lyap_k$ holds whenever $\nrm{\xh_k}\le\varrho_0(s)$ and the candidate $\widetilde d$ defined below is feasible at time $k+1$.
With $d=d_k^*$, the candidate is obtained by shifting $d$ by one stage, appending the local law at the predicted terminal state, and correcting the initial state to the measured state $\xh_{k+1}$, that is,
\begin{equation}\label{eq:candidate}
\widetilde d=\Ss d+\Ls\omega_k,\quad
\Ss=\col(\Th_1^x,J_1,\ldots,J_{N-1},K_c\Th_N^x),
\end{equation}
for which $J_z\widetilde d=\Th_1^xd+\omega_k=\xh_{k+1}$ since $J_z\Ls=I$.

We first bound the cost of the shifted vector $\Ss d$.
For the noise-free maps, the shift is exact: with $\Ss^0$ defined as $\Ss$ with $\Th_i^{x,0}$ in place of $\Th_i^x$, \eqref{eq:exactpred} gives $\Th_i^{x,0}\Ss^0=\Th_{i+1}^{x,0}$ for $i<N$ and $\Th_N^{x,0}\Ss^0=(A_z+B_zK_c)\Th_N^{x,0}$.
For the noisy maps, let $\Gamma_i=\Th_i^x\Ss-\Th_{i+1}^x$ for $i<N$ and $\Gamma_N=\Th_N^x\Ss-F\Th_N^x$, with $F$ of~\eqref{eq:cert-terminal}.
By~\eqref{eq:M}, $M$ consists of columns of $\Th_1^x$, so that~\eqref{eq:terr} also gives $\nrm{F-(A_z+B_zK_c)}\leq(1+\bar k)t$, so that subtracting the exact identities yields $\nrm{\Gamma_i}\le G_\Gamma t$ for some constant $G_\Gamma$.
Hence the state of $\Ss d$ at stage $i<N$ is $\Th_{i+1}^xd+\Gamma_id$, its terminal state is $F\Th_N^xd+\Gamma_Nd$, and its inputs are $J_1d,\ldots,J_{N-1}d$, and $K_c\Th_N^xd$.
The bound $\wnrm{p+q}{W}\le\wnrm{p}{W}+\nrm W(2\nrm p+\nrm q)\nrm q$ with $q=\Gamma_id$ therefore gives, for some constant $c_\Gamma$,
\begin{align}\label{eq:shiftcost}
(\Ss d)^\T\J\Ss d
&\le\sum_{i=1}^{N}\wnrm{\Th_i^xd}{Q}+\sum_{i=1}^{N-1}\wnrm{J_id}{R}
+\wnrm{K_c\Th_N^xd}{R}\nonumber\\
&\quad+\lambda\wnrm{F\Th_N^xd}{P}+c_\Gamma t\nrm d^2\nonumber\\
&\le d^\T\J d-\ell_c(J_zd,J_0d)+c_\Gamma t\nrm d^2,
\end{align}
where the second inequality holds because $C_K^\T C_K=Q+K_c^\T RK_c$ and~\eqref{eq:cert-terminal} give $\wnrm{\Th_N^xd}{Q}+\wnrm{K_c\Th_N^xd}{R}+\lambda\wnrm{F\Th_N^xd}{P}\leq\lambda\wnrm{\Th_N^xd}{P}$.

We next bound the storage function at the predicted state $\Th_1^xd$.
The storage condition~\eqref{eq:cert-storage} involves only $z_0=\col(J_zd,J_0d)$, whereas $\Th_1^xd$ also depends on $u_1,\ldots,u_{N-1}$ for noisy data.
The remainder $\pi_d=\Th_1^xd-Mz_0$ is the action of the corresponding columns of $\Th_1^x$, which vanish for $\Th_1^{x,0}$, so that $\nrm{\pi_d}\le t\nrm d$ by~\eqref{eq:terr}.
Let $x_b$ be a constant such that $\nrm{\Th_i^x}\leq x_b$ for every $i$.
Then $H_s\preceq h_+I$, $\nrm M\le x_b$, $t\le1$, and~\eqref{eq:cert-storage} at $z_0$ give, with $c_s=h_+(2x_b+1)$,
\begin{align}\label{eq:storage-t}
\wnrm{\Th_1^xd}{H_s}
&\le\wnrm{Mz_0}{H_s}+c_st\nrm d^2\nonumber\\
&\le\wnrm{J_zd}{H_s}-\varepsilon_s\nrm{J_zd}^2\nonumber\\
&\quad+\tfrac12\ell_c(J_zd,J_0d)+c_st\nrm d^2.
\end{align}

We now combine the two bounds.
Since $\widetilde d$ is feasible and $J_z\widetilde d=\xh_{k+1}$, we have $V(\xh_{k+1})\le\widetilde d^\T\J\widetilde d$, and hence $V_H(\xh_{k+1})\leq\widetilde d^\T\J\widetilde d+\wnrm{\Th_1^xd+\omega_k}{H_s}$.

Substituting $\widetilde d=\Ss d+\Ls\omega_k$ and using the bounds on $\nrm\Ss$, $\nrm\Ls$, $\nrm\J$, $\nrm{\Th_1^x}$, and $H_s$, we can bound the terms involving $\omega_k$. Combining these bounds with~\eqref{eq:shiftcost} and~\eqref{eq:storage-t}, and using $J_zd=\xh_k$, $J_0d=u_k$, and $d^\T\J d=V(\xh_k)$, we obtain that
\begin{align}\label{eq:robustpre}
V_H(\xh_{k+1})-V_H(\xh_k)\nonumber
&\le-\varepsilon_s\nrm{\xh_k}^2
    -\tfrac12\ell_c(\xh_k,u_k)+C_\Gamma t\nrm{d_k^*}^2\nonumber\\
&+C_\omega\nrm{d_k^*}\nrm{\omega_k}
    +C_\omega\nrm{\omega_k}^2,
\end{align}
for some constants $C_\Gamma$ and $C_\omega$.
By~\eqref{eq:local} and Young's inequality, $C_\omega\nrm{d_k^*}\nrm{\omega_k}\le\frac{\varepsilon_0}4\nrm{\xh_k}^2+\frac{C_\omega^2\nu^2}{\varepsilon_0}\nrm{\omega_k}^2$, and~\eqref{eq:noise} together with $(\sum_{j=1}^4p_j)^2\le4\sum_{j=1}^4p_j^2$ bounds $\nrm{\omega_k}^2$.
Let $D_\omega=C_\omega+C_\omega^2\nu^2/\varepsilon_0$, and let $t_*$ be small so  that $C_\Gamma\nu^2t_*+4D_\omega a_{\omega z}^2t_*^2\le\varepsilon_0/4$.
Then with $\varepsilon_s\ge\varepsilon_0$, $\nrm{d_k^*}\le\nu\nrm{\xh_k}$, and $\ell_c\ge0$, Eq.~\eqref{eq:robustpre} becomes
\begin{align}\label{eq:valueerror}
V_H(\xh_{k+1})-V_H(\xh_k)
&\le-\tfrac{\varepsilon_0}2\nrm{\xh_k}^2+4D_\omega\big(a_{\omega w}^2t^2\nrm{\xi_k}^2\nonumber\\
&\quad+E_v^2\nrm{\mathbf e_k}^2+E_e^2\nrm{e_k}^2\big).
\end{align}

The term in $\nrm{\xi_k}^2$ is compensated by the window term $\varphi_k$ of~\eqref{eq:W}.
Indeed, with $r_k=\col(u_k,y_k)$, shifting the sum gives the identity in~\eqref{eq:window}, and $\nrm{r_k}^2=\nrm{u_k}^2+\nrm{y_k}^2$, $\nrm{u_k}\le\nu\nrm{\xh_k}$, and~\eqref{eq:ybound} give the subsequent bound for some constants $q_z$, $q_v$, and $q_e$, provided that $t_*$ is so small that $4\Tini c_{yw}^2t_*^2\le\frac12$:
\begin{align}\label{eq:window}
\varphi_{k+1}-\varphi_k&=\Tini\nrm{r_k}^2-\nrm{\xi_k}^2\nonumber\\
&\le-\tfrac12\nrm{\xi_k}^2+q_z\nrm{\xh_k}^2\nonumber\\
&\quad+q_v\nrm{\mathbf e_k}^2+q_e\nrm{e_k}^2.
\end{align}
Let $\beta=\min\{1,\varepsilon_0/(4q_z)\}$, and let $t_*$ be so small that $16D_\omega a_{\omega w}^2t_*^2\le\beta$.
Adding $\beta$ times~\eqref{eq:window} to~\eqref{eq:valueerror} then gives, for some constants $D_v$ and $D_e$,
\begin{align*}
\Lyap_{k+1}-\Lyap_k
&\le-\tfrac{\varepsilon_0}4\nrm{\xh_k}^2-\tfrac\beta4\nrm{\xi_k}^2\\
&\quad+D_v\nrm{\mathbf e_k}^2+D_e\nrm{e_k}^2.
\end{align*}
Since $\Lyap_k\le c_+\nrm{\xh_k}^2+\beta\Tini\nrm{\xi_k}^2$ by~\eqref{eq:local} and $\varphi_k\le\Tini\nrm{\xi_k}^2$, the right-hand side is at most $-\theta\Lyap_k+D_v\nrm{\mathbf e_k}^2+D_e\nrm{e_k}^2$ with $\theta=\min\{\varepsilon_0/(4c_+),1/(4\Tini)\}\in(0,1)$.
With $\nrm{\mathbf e_k}^2\le\Tini\ee^2$, $\nrm{e_k}^2\le\ee^2$, and $D_*=\Tini D_v+D_e$, we obtain
\begin{equation}\label{eq:decrease}
\Lyap_{k+1}\le(1-\theta)\Lyap_k+D_*\ee^2.
\end{equation}
The tolerance $t_*$ is fixed as any number in $(0,t_1]$ that satisfies the three conditions imposed in this part.

\proofhead{Part III}
We finally establish recursive feasibility, constraint satisfaction, and the stability bound~\eqref{eq:iss}.
To this end, we first show that $\Lyap_k\le c(s)$ implies the two conditions of Part~II, the constraints at time $k$, and $\Lyap_{k+1}\le c(s)$, where
\begin{equation}\label{eq:c}
c(s)=\min\!\left\{
h_-\varrho_0(s)^2,\
\frac{\bar\delta(s)^2}{4B_{\rm sh}^2},\
\frac{s^2}{4t_*^2B_y^2}\right\}
\end{equation}
with the constants $B_{\rm sh}$ and $B_y$ introduced below.
Let $\Lyap_k\le c(s)$.
Then $V(\xh_k)<\infty$, so that~\eqref{eq:qp} is feasible at time $k$, $u_k=J_0d_k^*\in\U$, and $\Th_0^yd_k^*\in\Ys$.
Moreover, $V_H(\xh)\ge h_-\nrm\xh^2$, $\varphi_k\ge\nrm{\xi_k}^2$, and~\eqref{eq:local} give $\nrm{\xh_k}\le\sqrt{c(s)/h_-}\le\varrho_0(s)$, $\nrm{\xi_k}\le\sqrt{c(s)/\beta}$ and $\nrm{d_k^*}\le\sqrt{c(s)/h_0}$,
which is the first condition of Part~II.
Inserting these bounds and $\nrm{\mathbf e_k}\le\sqrt\Tini\,\ee$ into~\eqref{eq:noise} and into $\nrm{\widetilde d}\le\nrm\Ss\nrm{d_k^*}+L_b\nrm{\omega_k}$ gives, for some constants $B_{\rm sh}$, $B_y$, $E_\omega$, and $E_y$,
\begin{align}\label{eq:feasest}
\nrm{\widetilde d}&\le B_{\rm sh}\sqrt{c(s)}+L_bE_\omega\ee,\nonumber\\
\nrm{y_k-\Th_0^yd_k^*}&\le t_*B_y\sqrt{c(s)}+E_y\ee.
\end{align}
Let $\kappa=\sqrt{D_*/\theta}$, and choose the constants of the theorem as
\begin{equation}\label{eq:ab}
\begin{aligned}
a&=2\max\{E_y,\ t_*B_y\kappa\},\\
b&=2o_b\max\{L_bE_\omega,\ L_b\kappa/\sqrt{h_-},\ B_{\rm sh}\kappa\}.
\end{aligned}
\end{equation}
Since $o_b\ge1$, condition~\ref{it:noisy-margin}) gives $\bar\delta(s)\ge b\ee/o_b$ and $s>a\ee$, and hence, by~\eqref{eq:ab},
\begin{equation}\label{eq:margins}
L_bE_\omega\ee\le\frac{\bar\delta(s)}2,\quad E_y\ee<\frac s2,
\end{equation}
as well as
\begin{align*}
h_-\varrho_0(s)^2&=\frac{h_-\bar\delta(s)^2}{4L_b^2}\ge\kappa^2\ee^2,\\
\frac{\bar\delta(s)^2}{4B_{\rm sh}^2}&\ge\kappa^2\ee^2,\quad
\frac{s^2}{4t_*^2B_y^2}\ge\kappa^2\ee^2,
\end{align*}
so that $c(s)\ge\kappa^2\ee^2$.
By~\eqref{eq:c}, \eqref{eq:feasest}, and~\eqref{eq:margins}, we have $\nrm{\widetilde d}\le\bar\delta(s)$, so that $\widetilde d$ is feasible, which is the second condition of Part~II, and $\nrm{y_k-\Th_0^yd_k^*}<s$, so that $y_k\in\Th_0^yd_k^*+s\B\subseteq\Y$.
Hence~\eqref{eq:decrease} holds, and $D_*\ee^2=\theta\kappa^2\ee^2\le\theta c(s)$ gives $\Lyap_{k+1}\le c(s)$.

By induction from $\Lyap_0\le c(s)$, the closed loop is recursively feasible, satisfies $u_k\in\U$ and $y_k\in\Y$ for every $k\ge0$, and satisfies~\eqref{eq:decrease} at every $k$.
Iterating~\eqref{eq:decrease} gives $\Lyap_k\le(1-\theta)^k\Lyap_0+\kappa^2\ee^2$, and $h_-\nrm{\xh_k}^2\le\Lyap_k$ then yields~\eqref{eq:iss} with $\rho=1-\theta$, $\gamma_0=h_-^{-1/2}$, and $\gamma=\kappa h_-^{-1/2}$.

\end{proof}

\begin{remark}\label{rem:noise-free}
In Theorem~\ref{thm:noisy}, the effect of the noise scales with the noise level $\ee$.
As $\ee$ decreases, the neighborhood $\gamma\ee$ in~\eqref{eq:iss} shrinks to the origin and the margins $a\ee$ and $b\ee$ required in condition~\ref{it:noisy-margin}) vanish, so that, in the limit $\ee=0$, \eqref{eq:iss} gives exponential convergence of the data-driven state estimate to the origin, which is consistent with Theorem~\ref{thm:nominal}.
\end{remark}

\begin{remark}\label{rem:cert-noisy}
Condition~\ref{it:noisy-cert}) of Theorem~\ref{thm:noisy} can be verified directly from the data, since the certificate given by~\eqref{eq:term} and~\eqref{eq:cert} depends on the data only through the one-step predictor $M$ in~\eqref{eq:M}.
Hence Step~\ref{alg:cert} of Algorithm~\ref{alg:deopc} solves the corresponding LMIs with the predictor $M$ computed from the noisy data, and the condition holds whenever a solution is found.
To this end, a natural choice is $Q\succ0$, for which the certificate is feasible in the noise-free case by Lemma~\ref{lem:storage}.
\end{remark}

\section{Implementation}\label{sec:impl}
The weight $Q$ of DeoPC acts on the data-driven state estimate $\xh=Tx$, whose coordinates are determined by the data and differ in general from those of the internal state.
In this section, we present two ways to implement DeoPC in view of this fact.
If a few offline measurements of the state are available, the data-driven state is calibrated so that the weight and the reference are specified in physical coordinates, as shown in Section~\ref{sec:calibration}.
If no such measurements are available, DeoPC can be applied directly, and $Q$ is tuned in the same way as the state weight of MPC, as discussed in Section~\ref{sec:tuning}.

\subsection{Calibration From a Few Offline State Samples}\label{sec:calibration}

In some applications, the state can be measured offline at a few time instants, e.g., by laboratory analysis of samples, although it is not available online.
Suppose that $q\geq n$ such samples $x_{t_1},\ldots,x_{t_q}$ are available, each paired with the offline past window $\xi_{t_i}$, where the instants $t_i\geq\Tini$ are arbitrary and need not be consecutive.
We call the state in the coordinates of these samples the \emph{physical state} and, without loss of generality, take the plant~\eqref{eq:plant} in these coordinates.
The idea of the calibration is to estimate the transformation between the data-driven state estimate and the physical state from the pairs $(x_{t_i},\xh_{t_i})$ with $\xh_{t_i}=\Ez\xi_{t_i}$.
With $X=[x_{t_1}\ \cdots\ x_{t_q}]$ and $\Xh=[\xh_{t_1}\ \cdots\ \xh_{t_q}]$, the least-squares estimate is the calibration matrix $\widehat M=X\Xh^\dagger$.
A weight $Q_x\succeq0$ on the physical state and an equilibrium $(x^r,u^r)$ of the plant are then translated into the weight $Q=\widehat M^\T Q_x\widehat M$ and the reference $\xh^r=\widehat M^{-1}x^r$ for the data-driven state estimate, and the objective of~\eqref{eq:qp} is replaced by
\begin{equation}\label{eq:phy-cost}
\sum_{i=0}^{N-1}\Big(\wnrm{\xh_{i|k}-\xh^r}{Q}+\wnrm{u_{i|k}-u^r}{R}\Big)+\lambda\wnrm{\xh_{N|k}-\xh^r}{P},
\end{equation}
so that~\eqref{eq:qp} remains a quadratic program (QP).
The above procedure is summarized in  Algorithm~\ref{alg:align}.

\begin{algorithm}
\caption{DeoPC With a Few State Samples}\label{alg:align}
\begin{algorithmic}[1]
\Require the inputs of Algorithm~\ref{alg:deopc}; paired samples $(x_{t_i},\xi_{t_i})$, $i=1,\ldots,q$, $q\geq n$, at offline instants $t_i\geq\Tini$ in any order; weight $Q_x\succeq0$; equilibrium $(x^r,u^r)$.
\State Compute $\Ez$ as in Algorithm~\ref{alg:deopc}, $X=[x_{t_1}\ \cdots\ x_{t_q}]$, and $\Xh=[\xh_{t_1}\ \cdots\ \xh_{t_q}]$ with $\xh_{t_i}=\Ez\xi_{t_i}$.\label{alg:align-state}
\State If $\rank\Xh<n$ or $\sigma_n(\Xh)/\sigma_1(\Xh)$ is below a chosen threshold, add or replace samples and repeat Step~\ref{alg:align-state}.\label{alg:align-check}
\State Compute $\widehat M=X\Xh^\dagger$; if $\widehat M$ is singular, add or replace samples and return to Step~\ref{alg:align-state}.
\State Set $Q=\widehat M^\T Q_x\widehat M$ and $\xh^r=\widehat M^{-1}x^r$.
\State Run Algorithm~\ref{alg:deopc} with this $\Ez$ and this $Q$, and with the objective~\eqref{eq:phy-cost} in~\eqref{eq:qp}.
\end{algorithmic}
\end{algorithm}

Let $y^r=Cx^r+Du^r$ denote the output at the equilibrium $(x^r,u^r)$.
The following corollary of Theorem~\ref{thm:nominal} shows that, for noise-free data, DeoPC with the calibrated weight and reference regulates the physical state to $x^r$.

\begin{corollary}\label{cor:tracking}
Let the conditions of Theorem~\ref{thm:nominal} hold with $\xh_0-\xh^r$, $\U-u^r$, and $\Y-y^r$ in place of $\xh_0$, $\U$, and $\Y$, and suppose that the state samples are noise-free and that $\rank\Xh=n$.
Then the closed loop of~\eqref{eq:plant} and~\eqref{eq:qp} with the objective~\eqref{eq:phy-cost} is recursively feasible, satisfies $u_k\in\U$ and $y_k\in\Y$ for every $k\geq0$, and satisfies $\nrm{x_k-x^r}\leq c\,\rho^k\nrm{x_0-x^r}$ for every $k\geq0$ with some $c>0$ and $\rho\in(0,1)$.
\end{corollary}

\begin{proof}
By Lemma~\ref{lem:similarity}, we have $\Xh=TX$ and $\rank\Xh=n$. Then $\Xh\Xh^\dagger=I$ and we have $\widehat M=X\Xh^\dagger=T^{-1}$, $Q=T^{-\T}Q_xT^{-1}$, and $\xh^r=Tx^r$.
Since $(\xh^r,u^r)$ is an equilibrium of $(A_z,B_z)$ with output $y^r$, the tracking problem can be reformulated as the regulation problem~\eqref{eq:qp} in the error variables $\xh-\xh^r$, $u-u^r$, and $y-y^r$, with the constraint sets shifted to $\U-u^r$ and $\Y-y^r$. 
The stability certificate remains valid under this transformation, as it is independent of the constraint sets and the reference. Therefore, Theorem~\ref{thm:nominal} applies to the error state $\xh_k-\xh^r$. The claimed bound then follows from $x_k-x^r=T^{-1}(\xh_k-\xh^r)$.
\end{proof}

\begin{remark}\label{rem:reference}
The reference $(\xh^r,u^r)$ can be replaced online by that of another equilibrium without recomputing the certificate.
The guarantee of Corollary~\ref{cor:tracking} then holds from any instant at which $\xh_k-\xh^r$ lies in the region for the current reference.
\end{remark}

\subsection{Tuning Without State Samples}\label{sec:tuning}

When no state samples are available, DeoPC can be applied as in Algorithm~\ref{alg:deopc}, and $Q$ is chosen directly as a weight on the data-driven state estimate.
By Theorem~\ref{prop:equiv}, such a weight acts as the weight $T^\T QT$ on the internal state of the plant.
The following corollary shows that no design freedom is lost in this way, since every state-feedback MPC is obtained by a suitable choice of the weights of DeoPC.

\begin{corollary}\label{cor:tuning}
Let the conditions of Theorem~\ref{prop:equiv} hold.
Then, for every MPC~\eqref{eq:ideal} with weights $Q_x\succeq0$ and $P_x\succ0$, there exist weights $Q\succeq0$ and $P\succ0$ such that DeoPC~\eqref{eq:qp} is equivalent to this MPC in the sense of Theorem~\ref{prop:equiv}.
\end{corollary}
\begin{proof}
Since $T$ is nonsingular, the weights $Q=T^{-\T}Q_xT^{-1}\succeq0$ and $P=\lambda^{-1}T^{-\T}P_xT^{-1}\succ0$ satisfy $Q_x=T^\T QT$ and $P_x=\lambda T^\T PT$, and the claim follows from Theorem~\ref{prop:equiv}.
\end{proof}

Hence the behavior of any state-feedback MPC can be obtained by tuning the weights of DeoPC, although the correspondence between $Q$ and $Q_x$ is unknown without state samples.
The weight $Q$ is therefore tuned as in MPC: starting from a simple choice such as the identity matrix, which admits a certificate for noise-free data by Lemma~\ref{lem:storage}, $Q$ and $R$ are adjusted on the basis of the closed-loop response, and Step~\ref{alg:cert} of Algorithm~\ref{alg:deopc} recomputes the certificate for each choice.
Moreover, if the certificate is feasible for some $Q$, it remains feasible when $Q$ is increased to $Q+Q_+$ with $Q_+\succeq0$, since $\ker(Q+Q_+)\subseteq\ker Q$ preserves the detectability in Lemma~\ref{lem:storage}.

\section{Simulations}\label{sec:num}
In this section, numerical examples are presented to illustrate the theoretical results and to compare the proposed DeoPC with state-feedback MPC and DeePC.

We consider the linearized batch reactor of \cite{walsh2001scheduling}, whose continuous-time model is discretized by exact zero-order hold with a sampling period of $0.1\,\mathrm{s}$, which gives the plant matrices
\begin{equation}\label{eq:batch}
\begin{aligned}
&A=\begin{bmatrix}
1.178&0.001&0.512&-0.403\\
-0.051&0.662&-0.011&0.061\\
0.076&0.335&0.561&0.382\\
-0.001&0.335&0.089&0.849
\end{bmatrix},\\ 
&B=\begin{bmatrix}
0.004&-0.088\\
0.467&0.001\\
0.213&-0.235\\
0.213&-0.016
\end{bmatrix},
C=\begin{bmatrix}
1&0&1&-1\\
0&1&0&0
\end{bmatrix}, D=0.
\end{aligned}      
\end{equation}

The plant is open-loop unstable with the spectral radius $\rho(A)=1.220>1$, and it can be verified that $(A,B)$ is controllable and $(C,A)$ is observable with lag $\ell=2$.
We set $\Tini=4$, $N=10$, $R=0.2I_2$, and $\U=\Y=[-2,2]^2$. For DeoPC, we choose $\lambda=1$ and the certificate constants $\mu=\varepsilon_0=h_-=10^{-3}$, $h_+=10^3$, $\bar p=10^4$, and $\bar k=100$.
The offline data of length $\Td=400$ are recorded under the prestabilizing feedback $u_k=K_{\rm pre}y_k+\tilde u_k$ with $K_{\rm pre}=\begin{bsmallmatrix}0&-2\\5&\phantom{-}0\end{bsmallmatrix}$, where $\tilde u_k$ is uniformly distributed on $[-1,1]^2$.
For the noise-free offline data, the input Hankel matrix of depth $\Tini+N+n=18$ has full row rank with smallest singular value $1.91$, so that $u^{\rm d}$ is persistently exciting of order $18$.
Throughout this section, a superscript $j$ denotes the $j$th component of a vector, e.g., $x^j$, $u^j$, and $x^{r,j}$.

\subsection{Equivalence With State-Feedback MPC}\label{sec:sim-equiv}

\begin{figure}[!htbp]
\centering
\includegraphics[width=.95\columnwidth]{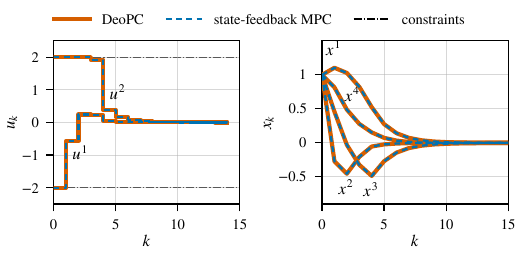}
\caption{Inputs (left) and physical states (right) of DeoPC (solid) and the state-feedback MPC (dashed) on noise-free data over the first $15$ steps, from $x_0=(1,1,1,1)^\T$. Dash-dotted lines mark the input constraints.}
\label{fig:equiv}
\end{figure}

We first consider the noise-free case and compare DeoPC with the state-feedback MPC~\eqref{eq:ideal}.
For DeoPC, we choose $Q=10I_4$ and $s=0$, and the weights of the MPC are set to $Q_x=T^\T QT$ and $P_x=\lambda T^\T PT$ as in Theorem~\ref{prop:equiv}.
Both closed loops start from $x_0=(1,1,1,1)^\T$, with an initial window of zero inputs and the corresponding outputs, and are simulated for $80$ steps.
As shown in Fig.~\ref{fig:equiv}, the trajectories of the two closed loops overlap, while the constraints are satisfied and every QP of both loops is feasible.
More precisely, the largest differences between the two loops are $2.3\times10^{-13}$ in the input, $7.5\times10^{-14}$ in the state, and $1.7\times10^{-12}$ in the relative difference of the optimal value.
Hence, the two closed loops coincide up to numerical errors, which is consistent with Theorem~\ref{prop:equiv}.

\subsection{Stability With Noisy Data}\label{sec:sim-noisy}

\begin{figure}[!htbp]
\centering
\includegraphics[width=.95\columnwidth]{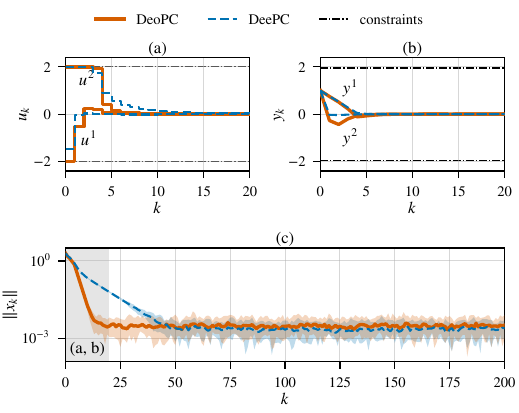}
\caption{DeoPC (solid) and tuned DeePC (dashed) on noisy data: (a) inputs and (b) outputs for $k=0,\ldots,20$ in the realization with the median DeoPC state energy, with dash-dotted input constraints and tightened output constraints. (c) Norm of the physical state over $200$ steps: median (lines) and range (shaded bands) over the $20$ realizations; the gray strip marks the steps of (a) and (b).}
\label{fig:noisy}
\end{figure}

We next consider the case of noisy data, in which the process noise and the measurement noise are drawn uniformly from Euclidean balls of radius $\bar w=\bar v=10^{-3}$, both offline and online.
For DeoPC, we choose $Q=10I_4$ and $s=0.05$ and simulate $20$ noise realizations of $200$ steps each from $x_0=(1,1,1,1)^\T$.
For comparison, we also consider DeePC~\eqref{eq:deepc} with $S=10I_2$, which by Remark~\ref{rem:deepc} can be viewed as a special case of DeoPC whose weight acts only on the outputs, and implement it with the regularization of~\cite{dorfler2023bridging,coulson2019data}, whose weights are tuned on a grid, and with the same $R$ and horizons. The results are shown in Fig.~\ref{fig:noisy}.

As shown in Fig.~\ref{fig:noisy}, the state of the DeoPC closed loop converges to a neighborhood of the origin with $\nrm{x_k}\leq8.1\times10^{-3}$ over the last $20$ steps in all realizations.
Moreover, in all realizations, the certificate is feasible, every QP is feasible, and the constraints are satisfied, which is consistent with Theorem~\ref{thm:noisy}.

\subsection{Tracking a Physical-State Reference}\label{sec:sim-tracking}

Finally, we consider the case where a few offline state samples are available (as discussed in Section~\ref{sec:impl}) and examine the performance of DeoPC in tracking a nonzero reference for the physical state.
Based on the noise-free data of Section~\ref{sec:sim-equiv}, we apply Algorithm~\ref{alg:align} with four state samples taken at the offline instants $t=4,100,196,292$, which yields $\widehat M$ with a calibration error of $\nrm{\widehat M-T^{-1}}/\nrm{T^{-1}}=7.0\times10^{-14}$.
The control objective is to regulate the states $x^3$ and $x^4$ to the targets $(x^3,x^4)=(0.8,0.8)$, $(0.3,0.5)$, and $(-0.6,-0.4)$, each of which is held for $20$ steps and determines an equilibrium $(x^r,u^r)$.
We also note that $x^3$ and $x^4$ are not measured directly, since $y=(x^1+x^3-x^4,x^2)$.
To this end, we choose $Q_x=10C^\T C+\diag(0,0,1,1)$ and $s=0$ for DeoPC, so that $x^3$ and $x^4$ are penalized in addition to the outputs.
For comparison, we also apply DeePC~\eqref{eq:deepc} with $S=10I_2$ to track the corresponding references $y^r=Cx^r$ and $u^r$, and both closed loops start from $x_0=0$.

\begin{figure}[!htbp]
\centering
\includegraphics[width=.95\columnwidth]{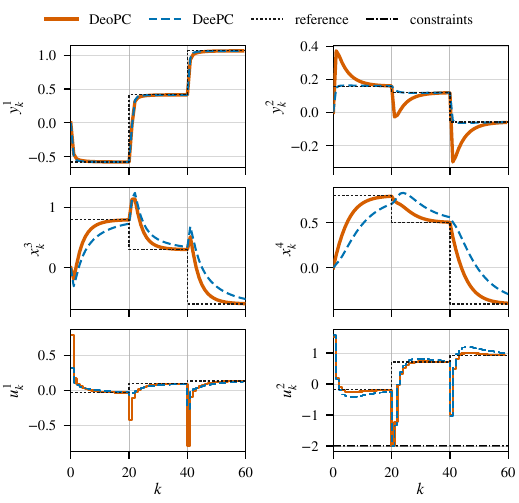}
\caption{Calibrated DeoPC (solid) and DeePC (dashed): outputs (top), $x^3,x^4$ (middle), and inputs (bottom), with the reference short-dashed, the constraint dash-dotted, and the switches at $k=20,40$.}
\label{fig:tracking}
\end{figure}

The closed-loop trajectories are shown in Fig.~\ref{fig:tracking}. All QPs are feasible and the constraints are satisfied.
It can be seen that both controllers track the output reference $y^r$ with DeePC responding faster.
For the states $x^3$ and $x^4$, however, DeoPC converges faster than DeePC to the targets.
This is expected, since DeePC penalizes only the outputs, whereas DeoPC can penalize $x^3$ and $x^4$ directly through the state weight.

\begin{figure}[!htbp]
\centering
\includegraphics[width=.95\columnwidth]{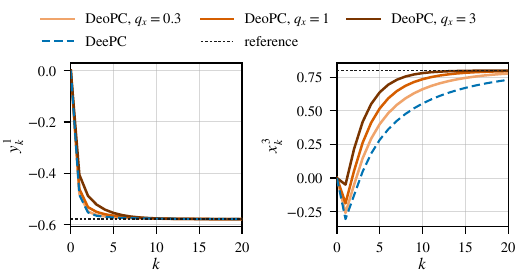}
\caption{Calibrated DeoPC with $q_x=0.3$, $1$, and $3$ (solid, darker for larger $q_x$) and DeePC (dashed) over the first $20$ steps: output $y^1$ (left) and state $x^3$ (right), with the reference short-dashed.}
\label{fig:tuning}
\end{figure}

To further illustrate the effect of the state weight, we repeat the above simulation with $Q_x=10C^\T C+q_x\diag(0,0,1,1)$ for $q_x=0.3$, $1$, and $3$, and the responses of $y^1$ and $x^3$ over the first $20$ steps are shown in Fig.~\ref{fig:tuning}.
It can be seen that, as $q_x$ increases, $x^3$ converges faster to its target at the price of a slightly slower output response.
Note that the stage cost of DeoPC with $q_x=0$ coincides with that of DeePC, so that $q_x$ directly adjusts the trade-off between the outputs and the internal states.

\section{Conclusions}\label{sec:concl}
In this work, we propose DeoPC, a predictive control framework for constrained state regulation of unknown linear systems using only input--output data.
Based on a data-driven state estimate constructed from past input--output measurements, DeoPC penalizes the internal state directly in the cost, and we show that it is equivalent to state-feedback MPC for noise-free data.
By certifying a storage function from the data, we establish recursive feasibility and exponential stability of the closed loop in the noise-free case.
For bounded noise, we further prove that the closed loop satisfies the constraints and is practically exponentially stable when the data are sufficiently exciting.
For implementation, we show that DeoPC can be calibrated to the physical state coordinates with a few offline state samples, or its state weight can be tuned to achieve state-feedback MPC when no state samples are available.
Finally, simulations on an unstable batch reactor validate the theoretical results.
Future work includes the extension of DeoPC to nonlinear systems and to stochastic noise.

\appendix
\subsection{Proof of Lemma~\ref{lem:pert}}
\label{app:perturbation}

\begin{proof}
Throughout the proof, a superscript $0$ marks the matrices that are built, as $E_0$ in~\eqref{eq:pert}, from the noise-free data, $\nrm{\cdot}_F$ denotes the Frobenius norm, and all constants depend only on the plant, on the fixed SVD of the noise-free data, and on $\Tini$, $N$, and $\Td$.

We first bound the perturbation of the data.
The noisy and the noise-free state differ by $\sum_{j<k}A^{k-1-j}w_j$, and~\eqref{eq:ebar} gives $\bar w\leq\bar\varepsilon\leq\ee/\nrm C$ and $\bar v\leq\ee$, so that every output sample differs from its noise-free counterpart by at most a constant multiple of $\ee$.
Hence there exist constants $c_d$ and $c_w$ such that $\nrm{W_i-W_i^0}\leq c_d\ee$, $\nrm{\Yf-\Yf^0}\leq c_d\ee$, and $\nrm{W_i}\leq c_w\dd$ for every $i$, whereas $\Uf$ and $\Pi^\perp$ do not depend on the noise.

We next bound the noise-free data matrices from below.
Let $X_s^0$ collect the noise-free states at the starting instants of the columns of $H_{\Tini+N}(u^{\rm d})$, and let $Z=\col(X_s^0,H_{\Tini+N}(u^{\rm d}))$.
By the quantitative state--input bound \cite[Thm.~3]{coulson2023quantitative}, $\smin(Z)\geq c_Z\sigma_d$ for some constant $c_Z$.
Moreover, $\col(\Wp^0,\Uf)=G_WZ$ and $\calA^0=G_AZ$ with fixed matrices $G_W$ and $G_A$, where $G_W$ has full column rank by observability and $\Tini\geq\ell$, and $G_A$ has full row rank by controllability.
Hence, with $a_\sigma=\smin(G_W)c_Z$ and $b_\sigma=\smin(G_A)c_Z$,
\[
\sigma_{m(\Tini+N)+n}\!\bigl(\col(\Wp^0,\Uf)\bigr)
\ge a_\sigma\sigma_d,\quad
\smin(\calA^0)\ge b_\sigma\sigma_d.
\]
Every $g\in\ker\Uf$ that is orthogonal to the kernel of $\col(\Wp^0,\Uf)$ satisfies $\nrm{\Wp^0g}=\nrm{\col(\Wp^0,\Uf)g}\geq a_\sigma\sigma_d\nrm g$, and these $g$ form a subspace of dimension $r_w$.
Hence $\Psi^0=\Wp^0\Pi^\perp$ has rank $r_w$ and $\sigma_{r_w}(\Psi^0)\geq a_\sigma\sigma_d$.

We now bound the perturbation of $L_w$.
Let $\Psi=\Wp\Pi^\perp$, let $\Psi_r$ denote its truncation to rank $r_w$, and suppose that $c_d\ee\leq a_\sigma\sigma_d/4$.
Since $\nrm{\Psi-\Psi^0}\leq c_d\ee$, Weyl's inequality gives $\sigma_{r_w}(\Psi)\geq3a_\sigma\sigma_d/4$ and $\sigma_{r_w+1}(\Psi)\leq a_\sigma\sigma_d/4$, so that the truncation in~\eqref{eq:Lw} is unique, $\nrm{\Psi_r^\dagger}\leq2/(a_\sigma\sigma_d)$, and $\nrm{\Psi_r-\Psi^0}\leq2c_d\ee$.
By the proof of Lemma~\ref{lem:similarity}, $L_w^0=\mathcal O_NF_p$ does not depend on the data and satisfies $\Yf^0\Pi^\perp=L_w^0\Psi^0$ and $L_w^0=L_w^0\mathcal P^0$ with the projector $\mathcal P^0=\Psi^0\Psi^{0\dagger}$.
With $\mathcal P_r=\Psi_r\Psi_r^\dagger$ and $L_w=\Yf\Pi^\perp \Psi_r^\dagger$, this gives
\[
L_w-L_w^0=\big(\Yf\Pi^\perp-L_w^0\Psi_r\big)\Psi_r^\dagger-L_w^0\mathcal P^0(I-\mathcal P_r).
\]
Here $\Yf\Pi^\perp-L_w^0\Psi_r=(\Yf-\Yf^0)\Pi^\perp+L_w^0(\Psi^0-\Psi_r)$ has norm at most $(1+2\nrm{L_w^0})c_d\ee$, and $\mathcal P^0(I-\mathcal P_r)=(\Psi^{0\dagger})^\T(\Psi^0-\Psi_r)^\T(I-\mathcal P_r)$ has norm at most $2c_d\ee/(a_\sigma\sigma_d)$, since $\Psi_r^\T(I-\mathcal P_r)=0$.
Hence $\nrm{L_w-L_w^0}\leq c_L\ee/\sigma_d$ with $c_L=2c_d(1+3\nrm{L_w^0})/a_\sigma$.

We now construct the orthogonal matrix $\Lambda$.
Let $s_0=\sigma_n(L_w^0)>0$, and suppose that $c_L\ee/\sigma_d\leq s_0/4$.
Since $L_w^0$ has rank $n$, Weyl's inequality gives $\sigma_n(L_w)\geq3s_0/4$ and $\sigma_{n+1}(L_w)\leq s_0/4$, so that the truncation in~\eqref{eq:Lw-svd} is unique and $\nrm{L_{w,n}-L_w^0}\leq2c_L\ee/\sigma_d$${}\leq s_0/2$.
The rank-$n$ matrices $L_{w,n}$ and $L_w^0$ therefore satisfy the conditions of~\cite[Lemma~5.14]{tu2015lowrank}, and their right factors $V_n\Sigma_n^{1/2}$ and $V_0\Sigma_0^{1/2}$ in this lemma are $\Ez^\T$ for every choice of the SVD in~\eqref{eq:Lw-svd} and $E_0^\T$ for the fixed SVD $L_w^0=U_0\Sigma_0V_0^\T$.
Hence there exists an orthogonal matrix $\Lambda$ such that $\nrm{\Ez-\Lambda E_0}_F^2\leq\frac{2}{\sqrt2-1}\nrm{L_{w,n}-L_w^0}_F^2/s_0$.
Since $L_{w,n}-L_w^0$ has rank at most $2n$, this gives $\nrm{\Ez-\Lambda E_0}\le C_E\ee/\sigma_d$ with $C_E=4c_L\sqrt{n/((\sqrt2-1)s_0)}$.

Finally, we bound the prediction maps.
Since $\Xh_i=\Ez W_i$ and $\Xh_i^0=E_0W_i^0$, we have $\Lambda^\T\Xh_i-\Xh_i^0=(\Lambda^\T\Ez-E_0)W_i+E_0(W_i-W_i^0)$, and hence $\nrm{\Lambda^\T\Xh_i-\Xh_i^0}\leq\delta_{\calA}$ with $\delta_{\calA}=c_0\ee+c_1\ee\dd/\sigma_d$, $c_0=\nrm{E_0}c_d$, and $c_1=C_Ec_w$.
The matrix $\widetilde{\calA}=\Delta_\Lambda^\T\calA=\col(\Lambda^\T\Xh_0,\Uf)$ therefore satisfies $\nrm{\widetilde{\calA}-\calA^0}\leq\delta_{\calA}$.
Suppose that $\delta_{\calA}\leq b_\sigma\sigma_d/2$.
Then $\smin(\widetilde{\calA})\geq b_\sigma\sigma_d/2$, so that $\calA$ has full row rank and $\nrm{\widetilde{\calA}^\dagger}\leq2/(b_\sigma\sigma_d)$.
Since $\Delta_\Lambda$ is orthogonal, $\calA^\dagger\Delta_\Lambda=\widetilde{\calA}^\dagger$, and since $\Th_i^{x,0}\calA^0=\Xh_i^0$ and $\Th_i^{y,0}\calA^0=Y_i^0$ by~\eqref{eq:exactpred}, we obtain
\begin{align*}
\Lambda^\T\Th_i^x\Delta_\Lambda-\Th_i^{x,0}
&=\big(\Lambda^\T\Xh_i-\Xh_i^0-\Th_i^{x,0}(\widetilde{\calA}-\calA^0)\big)\widetilde{\calA}^{\dagger},\\
\Th_i^y\Delta_\Lambda-\Th_i^{y,0}
&=\big(Y_i-Y_i^0-\Th_i^{y,0}(\widetilde{\calA}-\calA^0)\big)\widetilde{\calA}^{\dagger}.
\end{align*}
By~\eqref{eq:exactpred}, the norms of $\Th_i^{x,0}$ and $\Th_i^{y,0}$ are bounded by constants.
Together with $\nrm{Y_i-Y_i^0}\leq c_d\ee$ and the bounds on $\delta_{\calA}$ and $\nrm{\widetilde{\calA}^\dagger}$, the two identities and the bound on $\nrm{\Ez-\Lambda E_0}$ yield $t(\Lambda)\le C_1\ee/\sigma_d+C_2\ee\dd/\sigma_d^2$ for some constants $C_1$ and $C_2$.

The three conditions imposed above and the condition $C_1\ee/\sigma_d+C_2\ee\dd/\sigma_d^2<\tau$ bound quantities that are nonincreasing in $\sigma_d$ and invariant under a common scaling of $(\ee,\dd,\sigma_d)$.
Let $\sigma_\star(\ee,\dd;\tau)$ be the infimum of the values of $\sigma_d$ for which all four conditions hold.
Then they hold for every $\sigma_d>\sigma_\star(\ee,\dd;\tau)$, the threshold satisfies $\sigma_\star(r\ee,r\dd;\tau)=r\,\sigma_\star(\ee,\dd;\tau)$ for every $r>0$, and $\sigma_\star(0,\dd;\tau)=0$.
\end{proof}

\subsection{Proof of Lemma~\ref{lem:state-error}}
\label{app:state-error}

\begin{proof}
Consider the innovation form~\eqref{eq:innov} of a trajectory of~\eqref{eq:plant} that starts at a time $k_0\leq k-\Tini$, and let $\mathbf e_k=\col(e_{k-\Tini},\ldots,e_{k-1})$.
Let $\mathcal K_e=[A^{\Tini-1}K\ \cdots\ AK\ K]$, and let $\mathcal T_e$ be the block lower triangular Toeplitz matrix with blocks $I$ on the diagonal and $CA^{i-j-1}K$ in block $(i,j)$, $i>j$.
By~\eqref{eq:innov}, the noise-free response of~\eqref{eq:plant} from the state $\bar x_{k-\Tini}$ to the inputs $u_{k-\Tini},\ldots,u_{k-1}$ has the final state $\bar x_k-\mathcal K_e\mathbf e_k$ and the past window $\xi_k-\col(0,\mathcal T_e\mathbf e_k)$.
Since $E_0$ maps every noise-free past window to $T$ times its final state, it follows that $E_0(\xi_k-\col(0,\mathcal T_e\mathbf e_k))=T(\bar x_k-\mathcal K_e\mathbf e_k)$, and hence, with $\xh_k=\Ez\xi_k$,
\begin{equation}\label{eq:innovation}
\Lambda^\T\xh_k-T\bar x_k=(\Lambda^\T\Ez-E_0)\xi_k+R_v\mathbf e_k,
\end{equation}
where $R_v=E_0\col(0,\mathcal T_e)-T\mathcal K_e$.
Lemma~\ref{lem:innov} gives $\nrm{\mathbf e_k}\leq\sqrt{\Tini}\,\ee$ and $\nrm{x_k-\bar x_k}\leq\bar\varepsilon\leq\ee/\nrm C$, and Lemma~\ref{lem:pert} gives $\nrm{\Lambda^\T\Ez-E_0}\leq C_E\ee/\sigma_d$.
Hence~\eqref{eq:innovation} and the triangle inequality yield~\eqref{eq:state-error} with $c_x=\sqrt{\Tini}\nrm{R_v}+\nrm T/\nrm C$.
\end{proof}

\bibliographystyle{IEEEtran}
\bibliography{refs}

\begin{thebibliography}{10}
\providecommand{\url}[1]{#1}
\csname url@samestyle\endcsname
\providecommand{\newblock}{\relax}
\providecommand{\bibinfo}[2]{#2}
\providecommand{\BIBentrySTDinterwordspacing}{\spaceskip=0pt\relax}
\providecommand{\BIBentryALTinterwordstretchfactor}{4}
\providecommand{\BIBentryALTinterwordspacing}{\spaceskip=\fontdimen2\font plus
\BIBentryALTinterwordstretchfactor\fontdimen3\font minus
  \fontdimen4\font\relax}
\providecommand{\BIBforeignlanguage}[2]{{%
\expandafter\ifx\csname l@#1\endcsname\relax
\typeout{** WARNING: IEEEtran.bst: No hyphenation pattern has been}%
\typeout{** loaded for the language `#1'. Using the pattern for}%
\typeout{** the default language instead.}%
\else
\language=\csname l@#1\endcsname
\fi
#2}}
\providecommand{\BIBdecl}{\relax}
\BIBdecl

\bibitem{hou2013model}
Z.-S. Hou and Z.~Wang, ``From model-based control to data-driven control:
  Survey, classification and perspective,'' \emph{Information Sciences}, vol.
  235, pp. 3--35, 2013.

\bibitem{markovsky2021behavioral}
I.~Markovsky and F.~D\"orfler, ``Behavioral systems theory in data-driven
  analysis, signal processing, and control,'' \emph{Annual Reviews in Control},
  vol.~52, pp. 42--64, 2021.

\bibitem{coulson2019data}
J.~Coulson, J.~Lygeros, and F.~D\"orfler, ``Data-enabled predictive control: In
  the shallows of the {DeePC},'' in \emph{2019 18th European Control Conference
  (ECC)}, 2019, pp. 307--312.

\bibitem{elokda2021data}
E.~Elokda, J.~Coulson, P.~N. Beuchat, J.~Lygeros, and F.~D\"orfler,
  ``Data-enabled predictive control for quadcopters,'' \emph{International
  Journal of Robust and Nonlinear Control}, vol.~31, no.~18, pp. 8916--8936,
  2021.

\bibitem{huang2022decentralized}
L.~Huang, J.~Coulson, J.~Lygeros, and F.~D\"orfler, ``Decentralized
  data-enabled predictive control for power system oscillation damping,''
  \emph{IEEE Transactions on Control Systems Technology}, vol.~30, no.~3, pp.
  1065--1077, 2022.

\bibitem{carlet2022data}
P.~G. Carlet, A.~Favato, S.~Bolognani, and F.~D\"orfler, ``Data-driven
  continuous-set predictive current control for synchronous motor drives,''
  \emph{IEEE Transactions on Power Electronics}, vol.~37, no.~6, pp.
  6637--6646, 2022.

\bibitem{chinde2022data}
V.~Chinde, Y.~Lin, and M.~J. Ellis, ``Data-enabled predictive control for
  building {HVAC} systems,'' \emph{Journal of Dynamic Systems, Measurement, and
  Control}, vol. 144, no.~8, p. 081001, 2022.

\bibitem{joseph1978inferential}
B.~Joseph and C.~B. Brosilow, ``Inferential control of processes: {Part I}.
  {Steady} state analysis and design,'' \emph{AIChE Journal}, vol.~24, no.~3,
  pp. 485--492, 1978.

\bibitem{tsuji2026predictive}
K.~Tsuji, I.~Maruta, K.~Fujimoto, T.~Maeda, Y.~Tamase, and T.~Shinohara,
  ``Predictive virtual sensor identification for nonlinear {MPC}: application
  to a {2-DoF} flexible direct-drive robotic arm,'' \emph{SICE Journal of
  Control, Measurement, and System Integration}, vol.~19, no.~1, p. 2732446,
  2026.

\bibitem{depersis2020formulas}
C.~De~Persis and P.~Tesi, ``Formulas for data-driven control: Stabilization,
  optimality, and robustness,'' \emph{IEEE Transactions on Automatic Control},
  vol.~65, no.~3, pp. 909--924, 2020.

\bibitem{dorfler2023certainty}
F.~D\"orfler, P.~Tesi, and C.~De~Persis, ``On the certainty-equivalence
  approach to direct data-driven {LQR} design,'' \emph{IEEE Transactions on
  Automatic Control}, vol.~68, no.~12, pp. 7989--7996, 2023.

\bibitem{favoreel1999model}
W.~Favoreel, B.~De~Moor, P.~Van~Overschee, and M.~Gevers, ``Model-free
  subspace-based {LQG}-design,'' in \emph{Proceedings of the 1999 American
  Control Conference (Cat. No. 99CH36251)}, vol.~5, 1999, pp. 3372--3376.

\bibitem{dong2008equivalence}
J.~Dong and M.~Verhaegen, ``On the equivalence of closed-loop subspace
  predictive control with {LQG},'' in \emph{2008 47th IEEE Conference on
  Decision and Control (CDC)}, 2008, pp. 4085--4090.

\bibitem{dai2023data}
X.~Dai, C.~De~Persis, and N.~Monshizadeh, ``Data-driven optimal output feedback
  control of linear systems from input-output data,'' \emph{IFAC-PapersOnLine},
  vol.~56, no.~2, pp. 1376--1381, 2023.

\bibitem{alsalti2024efficient}
M.~Alsalti, V.~G. Lopez, and M.~A. M\"uller, ``An efficient data-based
  off-policy {Q}-learning algorithm for optimal output feedback control of
  linear systems,'' in \emph{Proceedings of the 6th Annual Learning for
  Dynamics \& Control Conference}, ser. Proceedings of Machine Learning
  Research, vol. 242, 2024, pp. 312--323.

\bibitem{zhang2011improved}
R.~Zhang, A.~Xue, S.~Wang, and Z.~Ren, ``An improved model predictive control
  approach based on extended non-minimal state space formulation,''
  \emph{Journal of Process Control}, vol.~21, no.~8, pp. 1183--1192, 2011.

\bibitem{zhang2018systematic}
R.~Zhang, S.~Wu, Z.~Cao, J.~Lu, and F.~Gao, ``A systematic min--max
  optimization design of constrained model predictive tracking control for
  industrial processes against uncertainty,'' \emph{IEEE Transactions on
  Control Systems Technology}, vol.~26, no.~6, pp. 2157--2164, 2018.

\bibitem{wang2022modeling}
J.~Wang, B.~Ding, and P.~Wang, ``Modeling and finite-horizon {MPC} for a
  boiler-turbine system using minimal realization state-space model,''
  \emph{Energies}, vol.~15, no.~21, p. 7935, 2022.

\bibitem{balim2024data}
H.~Balim, A.~Carron, M.~N. Zeilinger, and J.~K\"ohler, ``From data to
  predictive control: A framework for stochastic linear systems with output
  measurements,'' arXiv:2407.17277, 2024.

\bibitem{berberich2021data}
J.~Berberich, J.~K\"ohler, M.~A. M\"uller, and F.~Allg\"ower, ``Data-driven
  model predictive control with stability and robustness guarantees,''
  \emph{IEEE Transactions on Automatic Control}, vol.~66, no.~4, pp.
  1702--1717, 2021.

\bibitem{berberich2021terminal}
------, ``On the design of terminal ingredients for data-driven {MPC},''
  \emph{IFAC-PapersOnLine}, vol.~54, no.~6, pp. 257--263, 2021.

\bibitem{yang2026unified}
Z.~Yang, N.~Bai, P.~Wang, and L.~Shi, ``A unified framework for robust
  data-driven model predictive control with terminal ingredients,''
  arXiv:2610.04900, 2026.

\bibitem{bongard2023robust}
J.~Bongard, J.~Berberich, J.~K\"ohler, and F.~Allg\"ower, ``Robust stability
  analysis of a simple data-driven model predictive control approach,''
  \emph{IEEE Transactions on Automatic Control}, vol.~68, no.~5, pp.
  2625--2637, 2023.

\bibitem{alsalti2026robust}
M.~Alsalti, M.~Barkey, V.~G. Lopez, and M.~A. M\"uller, ``Robust and efficient
  data-driven predictive control,'' \emph{Automatica}, vol. 191, 2026, art. no.
  113108.

\bibitem{kloppelt2025novel}
C.~Kl\"oppelt, J.~Berberich, F.~Allg\"ower, and M.~A. M\"uller, ``A novel
  constraint-tightening approach for robust data-driven predictive control,''
  \emph{International Journal of Robust and Nonlinear Control}, vol.~35, no.~7,
  pp. 2566--2587, 2025.

\bibitem{favoreel1999spc}
W.~Favoreel, B.~De~Moor, and M.~Gevers, ``{SPC}: Subspace predictive control,''
  \emph{IFAC Proceedings Volumes}, vol.~32, no.~2, pp. 4004--4009, 1999.

\bibitem{vanoverschee1996subspace}
P.~Van~Overschee and B.~De~Moor, \emph{Subspace Identification for Linear
  Systems: Theory---Implementation---Applications}.\hskip 1em plus 0.5em minus
  0.4em\relax Kluwer Academic Publishers, 1996.

\bibitem{rawlings2017model}
J.~B. Rawlings, D.~Q. Mayne, and M.~M. Diehl, \emph{Model Predictive Control:
  Theory, Computation, and Design}, 2nd~ed.\hskip 1em plus 0.5em minus
  0.4em\relax Santa Barbara, CA: Nob Hill Publishing, 2017.

\bibitem{coulson2023quantitative}
J.~Coulson, H.~J. van Waarde, J.~Lygeros, and F.~D\"orfler, ``A quantitative
  notion of persistency of excitation and the robust fundamental lemma,''
  \emph{IEEE Control Systems Letters}, vol.~7, pp. 1243--1248, 2023.

\bibitem{berberich2023quantitative}
J.~Berberich, A.~Iannelli, A.~Padoan, J.~Coulson, F.~D\"orfler, and
  F.~Allg\"ower, ``A quantitative and constructive proof of {Willems'}
  fundamental lemma and its implications,'' in \emph{2023 American Control
  Conference (ACC)}, 2023, pp. 4155--4160.

\bibitem{willems2005note}
J.~C. Willems, P.~Rapisarda, I.~Markovsky, and B.~L.~M. De~Moor, ``A note on
  persistency of excitation,'' \emph{Systems \& Control Letters}, vol.~54,
  no.~4, pp. 325--329, 2005.

\bibitem{fiedler2021relationship}
F.~Fiedler and S.~Lucia, ``On the relationship between data-enabled predictive
  control and subspace predictive control,'' in \emph{2021 European Control
  Conference (ECC)}, 2021, pp. 222--229.

\bibitem{dorfler2023bridging}
F.~D\"orfler, J.~Coulson, and I.~Markovsky, ``Bridging direct and indirect
  data-driven control formulations via regularizations and relaxations,''
  \emph{IEEE Transactions on Automatic Control}, vol.~68, no.~2, pp. 883--897,
  2023.

\bibitem{verheijen2023handbook}
P.~C.~N. Verheijen, V.~Breschi, and M.~Lazar, ``Handbook of linear data-driven
  predictive control: Theory, implementation and design,'' \emph{Annual Reviews
  in Control}, vol.~56, p. 100914, 2023.

\bibitem{limon2006stability}
D.~Lim{\'o}n, T.~{\'A}lamo, F.~Salas, and E.~F. Camacho, ``On the stability of
  constrained {MPC} without terminal constraint,'' \emph{IEEE Transactions on
  Automatic Control}, vol.~51, no.~5, pp. 832--836, 2006.

\bibitem{walsh2001scheduling}
G.~C. Walsh and H.~Ye, ``Scheduling of networked control systems,'' \emph{IEEE
  Control Systems Magazine}, vol.~21, no.~1, pp. 57--65, 2001.

\bibitem{tu2015lowrank}
S.~Tu, R.~Boczar, M.~Simchowitz, M.~Soltanolkotabi, and B.~Recht, ``Low-rank
  solutions of linear matrix equations via {Procrustes} flow,''
  arXiv:1507.03566, 2015.

\end{thebibliography}

\PaperBios

\end{document}